\documentclass[a4paper,fleqn,usenatbib]{mnras} 
\usepackage{graphicx}
\usepackage{psfrag}
\usepackage{amsmath}
\usepackage{amssymb}
\usepackage{color}
\usepackage{hyperref}

\usepackage{subfigure,float}
\usepackage{multirow}
\usepackage[normalem]{ulem}

\usepackage{array,booktabs}
\usepackage{siunitx}

\newcommand{\beq}{\begin{equation}}
\newcommand{\eeq}{\end{equation}}

\newcommand{\ber}{\begin{eqnarray}}
\newcommand{\eer}{\end{eqnarray}}

\def\rhi {\rho_{HI}}

\def \meanrh {\bar{\rho}_H}
\def \ffc {FF_{\rm ionized}}

\def \meanrhi {\bar{\rho}_{HI}}

\def \xhi {x_{HI}}

\def\aj{AJ}
\def\apj{ApJ}
\def\apjl{ApJL}
\def\mnras{MNRAS}
\def\prd{PRD}

\def\aap{A\&A}
\def\apjs{ApJS}

\title[The Shape and Size distribution of HII Regions near the percolation transition]{The Shape and Size distribution of HII Regions near the percolation transition}

\author[Bag, Mondal, Sarkar, Bharadwaj, Sahni]{Satadru Bag,$^1$\thanks{E-mail: satadru@iucaa.in} Rajesh Mondal,$^{2,\,3}$ Prakash Sarkar,$^4$ 
\newauthor
Somnath Bharadwaj,$^3$  and Varun Sahni$^1$
\\
$^{1}$ Inter-University Centre for Astronomy and Astrophysics, Pune, India\\
$^{2}$ Astronomy Centre, Department of Physics and Astronomy, University of Sussex, Brighton, BN19QH, UK\\
$^{3}$ Department of Physics and Centre for Theoretical Studies, Indian Institute of Technology Kharagpur, Kharagpur 721302, India\\
$^{4}$ National Institute of Technology, Jamshedpur, India
}

\date{}

\pubyear{2015}

\begin{document}
\label{firstpage}
\pagerange{\pageref{firstpage}--\pageref{lastpage}}
\maketitle

\begin{abstract}
Using Shapefinders, which are ratios of Minkowski functionals, we study the morphology of neutral hydrogen (HI) density fields, simulated using semi-numerical technique (inside-out), at various stages of reionization. Accompanying the Shapefinders, we also 
employ the `largest cluster statistic' (LCS), originally proposed in \cite{Klypin1993},
 to study the percolation in both neutral and ionized 
hydrogen. We find that the largest ionized region is percolating below 
the neutral fraction $\xhi \lesssim 0.728$ (or equivalently $z \lesssim 9$).
The study of Shapefinders reveals that the largest ionized region starts to become highly filamentary with non-trivial topology near the percolation transition. During the percolation transition, the first two Shapefinders -- `thickness'
($T$) and `breadth' ($B$) -- of the largest ionized region do not vary 
much, while 
the third Shapefinder -- `length' ($L$) -- abruptly increases.
 Consequently, the largest ionized region tends 
to be highly filamentary and topologically quite complex.
The product of the first two Shapefinders, $T\times B$, provides a measure
of the `cross-section' of a filament-like ionized region. We find that,
near percolation, the value of $T\times B$ for the largest ionized region
remains stable at $\sim 7$ Mpc$^2$ (in comoving scale)
while its length increases with time.
Interestingly all large
ionized regions have similar cross-sections.
However, their length shows a power-law
 dependence on their volume, $L\propto V^{0.72}$, at the onset of percolation.
\end{abstract}

\begin{keywords}
intergalactic medium --  cosmology: theory --  dark ages, reionization, first stars -- large-scale structure of Universe
\end{keywords}

\section{Introduction}
It is widely believed that, following the cosmological recombination
of hydrogen at $z \simeq 1089$, the universe reionized at the much
lower redshift of $z \sim 10$ \citep{planck}. Our current knowledge
about this epoch of reionization (EoR) is guided so far by three main
observations. Measurements of the Thomson scattering optical depth
of CMB photons from free
electrons \citep{komatsu2011,planck2016a,planck2016b}, observations of the Lyman-$\alpha$ 
absorption spectra of the high-redshift quasars \citep{becker2001,
fan2003, goto2011, becker2015} and the luminosity function and
clustering properties of Lyman-$\alpha$ emitters \citep{trenti2010,
ouchi2010, ota2017, zheng2017}. These observations, when taken together, suggest
that the epoch of reionization probably extended over a wide redshift range
$6 \lesssim z \lesssim 15$ \citep{mitra2015, robertson2015}. Although
the precise physical mechanism responsible for cosmological
reionization is not known, it is widely believed that early sources of
energetic photons contributing to reionization may have come from: an
early generation of stars (population III objects), galaxies and
quasars \citep{loeb}. (More exotic sources such as decaying dark matter
have also been explored.)

Observation of the redshifted 21-cm signal from neutral hydrogen (HI)
provides an excellent means of studying the epoch of reionization and
the preceding `dark ages'. Considerable efforts are presently underway
to detect the EoR 21-cm signal using ongoing and upcoming radio
interferometric experiments e.g. GMRT \citep{paciga2013},
LOFAR \citep{haarlen2013,yatawatta2013},
MWA \citep{bowman2013,dillon2014},
PAPER \citep{parsons2014,ali2015,jacobs2015},
SKA \citep{mellema2013,koopmans2015}, HERA \citep{furlanetto2009, DeBoer2017}. 
The importance of a precise
determination of the epoch of reionization, and the associated
geometry and dynamics of neutral (HI) and ionized (HII) hydrogen regions, cannot be overstated.
Such an advance would open a new window into the physics of the early
universe, shedding light on important issues including the physics of
structure formation, the nature of feedback from the first collapsed
objects, the nature of dark matter and perhaps even dark energy.

To explore this vibrant reionization landscape we borrow tools originally developed for
the understanding and analysis of the cosmic web, which is similarly rich in geometrical properties.
For this purpose we use percolation analysis\footnote{Percolation has been studied comprehensively in the context of mathematical and condensed matter physics \citep{essam,Isichenko,stauffer,saberi}} \citep{Shandarin1983, Klypin1993} in conjunction with 
the Shapefinders, which are introduced in \citet{Sahni:1998cr} as ratios of Minkowski functionals, to assess the morphology
of reionized HII regions at different redshifts. These geometrical tools therefore play a role which is complementary to that of traditional
 N-point correlation functions. Our main method of analysis shall be the computationally advanced version of the 
 SURFGEN algorithm \citep{Sheth:2002rf} which, in the context of
structure formation, provides a means of determining the geometrical and topological
properties of isodensity contours delineating superclusters and voids within the 
cosmic web \citep{Sheth2004,Shandarin:2003tx,Sheth2005}. In this paper we refine and adapt this algorithm to determine the
 shapes and sizes of HII regions at different redshifts. The physics underlying cosmological reionization is expected to be 
reflected in the geometry and morphology
of HI and HII regions. For instance, the topology and morphology of reionization would be different if it  were driven by a few quasars instead of 
numerous galaxies.

In recent times many efforts have been made to understand reionization from geometrical points of view including granulometry \citep{Kakiichi2017},  percolation analyses \citep{Iliev2006,Iliev2014,Furlanetto2016} and the Minkowski functionals \citep{Friedrich2011,yoshiura:2017,Kapahtia2017}.
Our work, presented in this first of a series of papers, would be unique and interesting for two following reasons. Firstly, Shapefinders provide a direct measure of the geometry as well as the shape of each individual region and complement the indirect methods of estimating the shapes (for example fitting ellipsoids by \cite{Furlanetto2016}). Secondly, our advanced algorithm SURFGEN2, models surfaces through triangulation and the accuracy in measuring the Minkowski functionals and Shapefinders is much better compared to the existing widely used methods, such as the Crofton's formula \citep{Crofton_1968} which is based on cell counting.   

Our paper is organized as follows. The simulation of the HI field is briefly described in section \ref{sec:data}. In section \ref{sec:cir}, the ionized hydrogen (HII) region is analyzed using percolation. The shape of the ionized regions are studied in section \ref{sec:shape}. Our conclusions are presented in section \ref{sec:conclusion}.

\section{Simulating the neutral hydrogen (HI) density field}
\label{sec:data}
We have generated the neutral hydrogen field using semi-numerical
simulations. Our semi-numerical method involves three main steps: (i)
generating the dark matter distribution 
at the desired redshift, (ii) identifying the location and mass of
collapsed dark matter halos within the simulation volume, (iii)
generating the neutral hydrogen map using an excursion set formalism
\citep{furlanetto2004}. The assumption here is that the hydrogen
exactly traces the underlying dark matter field and the dark matter
halos host the ionizing sources. We discuss our method in the
following paragraphs. 

We have used a particle-mesh (PM) $N$-body code to generate the
dark matter distribution 
at desired redshifts.
Our simulation volume is a
$\sim [215\,{\rm Mpc}]^3$ comoving box. We have run our simulation with a
$3072^3$ grid using $1536^3$ dark matter particles. The spatial
resolution is $0.07\,{\rm Mpc}$ which corresponds to a mass resolution
of $1.09\times 10^8\,M_{\odot}$.


In the next step, we use the friends-of-friends (FoF) algorithm 
to identify the location and mass of the collapsed halos in the
dark matter distribution. We have used a fixed linking length, which is $0.2$
times the mean inter-particle separation and require a halo to have at
least $10$ dark matter particles which corresponds to a minimum halo
mass of $1.09\times 10^9\,M_{\odot}$.

In the final step, we generate the ionization map and the HI
distribution using the homogeneous recombination scheme
of \citet{Choudhury 2009}. It is assumed that the number of ionizing
photons emitted by a source is proportional to the mass of the host
halo with the constant of proportionality being quantified through a
dimensionless parameter $n_{\rm ion}$. In addition to $n_{\rm ion}$,
the simulations have another free parameter $R_{\rm mfp}$, the mean
free path of the ionizing photons. The final ionized maps were
generated on a grid that is $8$ times coarser than the $N$-body
simulations, i.e. with a spatial resolution of $0.56\,{\rm Mpc}$  within the simulation volume $\sim [215\,{\rm Mpc}]^3$ in comoving scale. Our semi-numerical simulations closely
follow \citet{majumdar2014, mondal2015, mondal2016, mondal2017,
mondal2018} to generate the ionization field.

The redshift evolution of the neutral fraction (the fraction of hydrogen mass that is ionized; $\xhi (z)=\meanrhi(z)/\meanrh(z)$) during EoR is
largely unknown. It is constrained from the CMB anisotropy and
polarization measurements \citep{planck2016b} and observations of the
Lyman-$\alpha$ absorption spectra of the high-redshift quasars
\citep{fan2003, becker2015}. These constraints can be satisfied for a
wide range of ionization histories. Given the uncertainty of
reionization history, the values of reionization model parameters were
fixed at $n_{\rm ion}=23.21$ and $R_{\rm mfp}=20\,{\rm
Mpc}$ \citep{songaila2010} so as to achieve 50\% ionization by
$z=8$. 

\section{Percolation analysis}\label{sec:cir}

\begin{figure}
\centering
\includegraphics[width=0.4\textwidth]{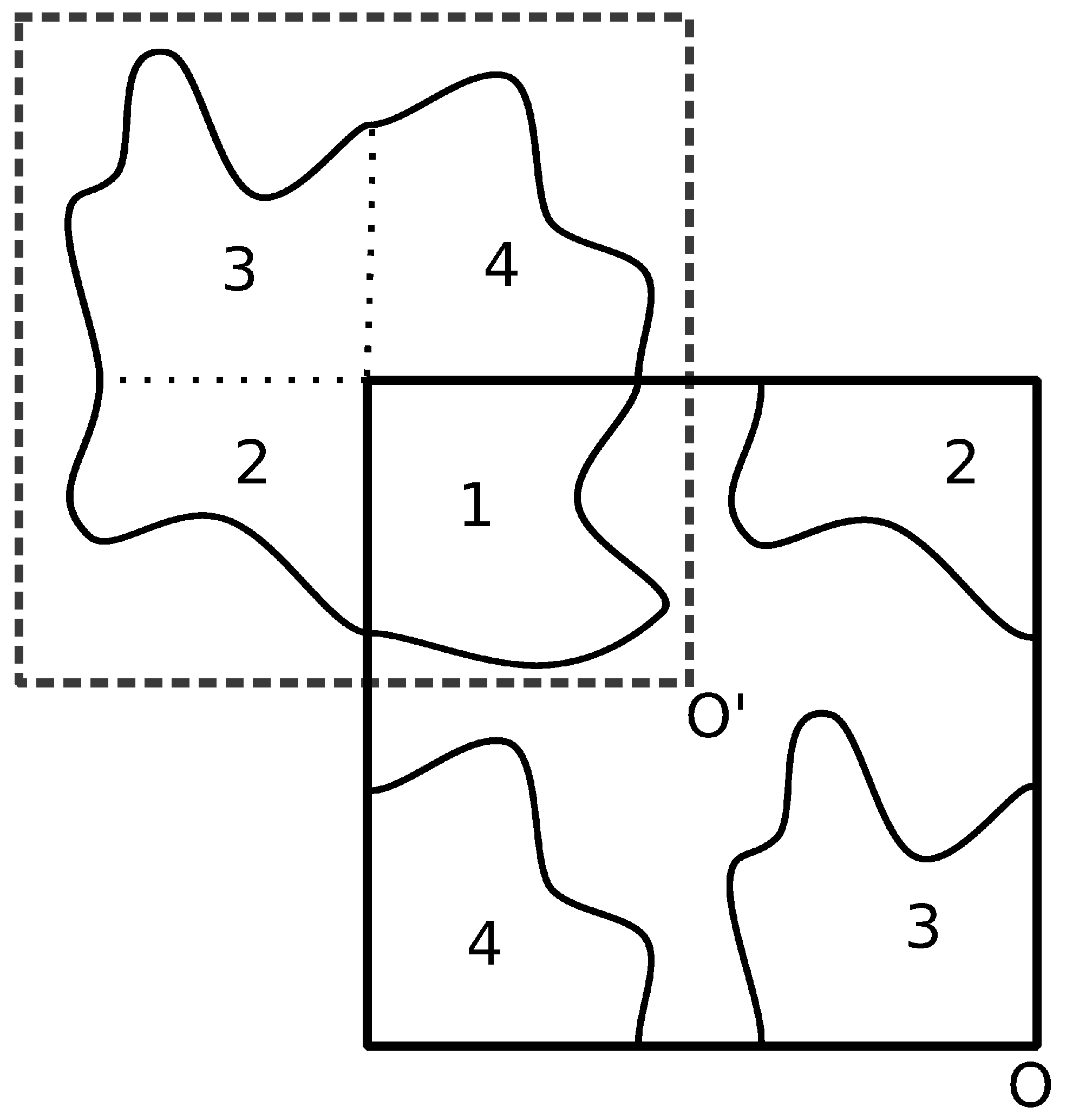}
\caption{The implementation of periodic boundary conditions (PBC) is explained using
 a two-dimensional representation. We define individual regions with connected grid points of same type (neutral or ionized) using friends-of-friends (FoF) algorithm.
Due to PBC
the (apparently) disconnected regions appearing the primary box, centered at $O$, are actually 
4 fragments of one single region.  
To model the surface of the whole region correctly, while calculating its Minkowski functionals and Shapefinders in section \ref{sec:shape}, 
we restore the whole region to its original 
shape by simply moving the box to its new origin at $O'$. 
This procedure can be easily generalized to 3D.}
\label{fig:pbc}
\end{figure}

\begin{figure*}[hb]
\centering
\subfigure[]{
\includegraphics[width=0.47\textwidth]{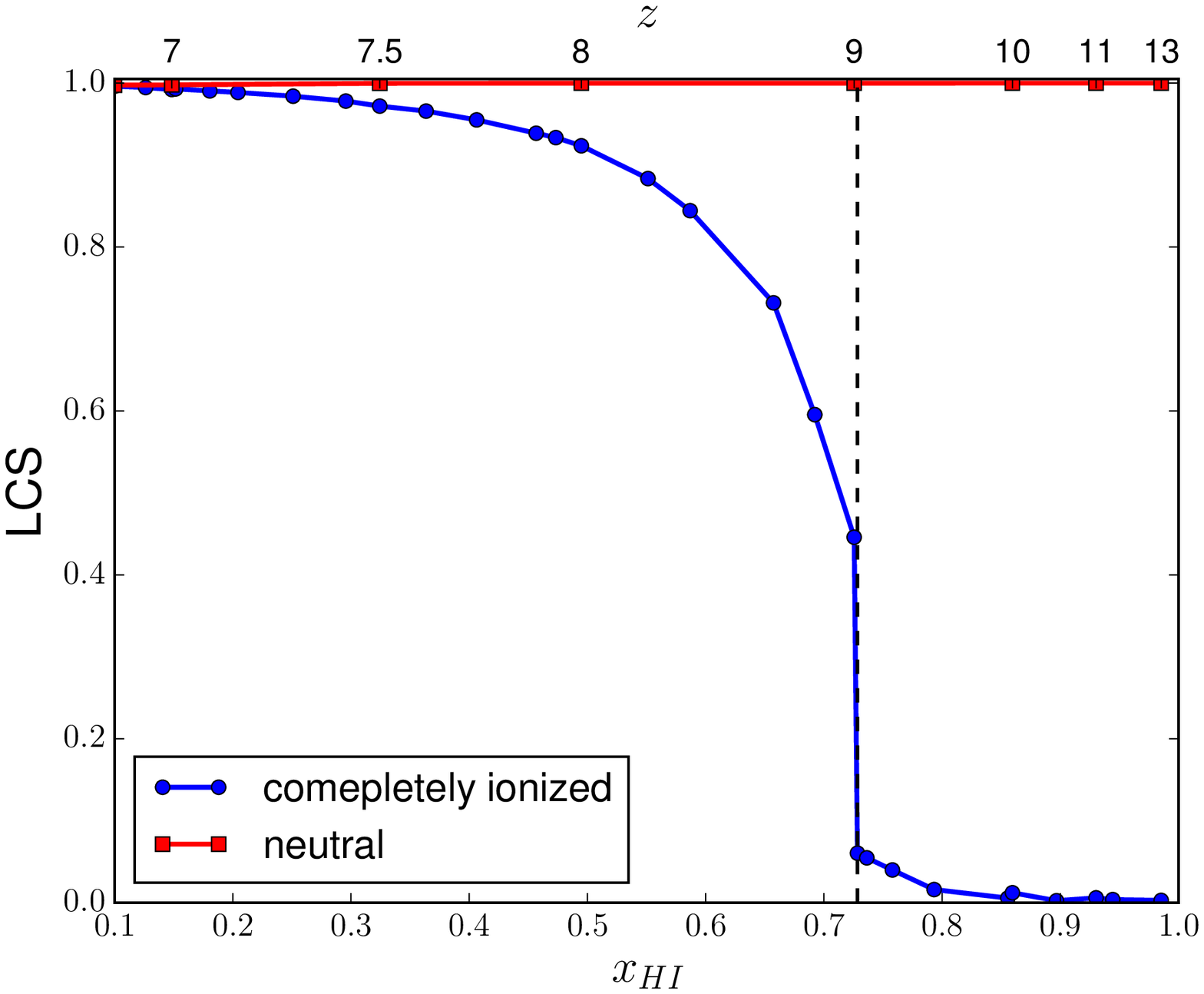}\label{fig:lcs_vs_xhi}}
\subfigure[]{
\includegraphics[width=0.47\textwidth]{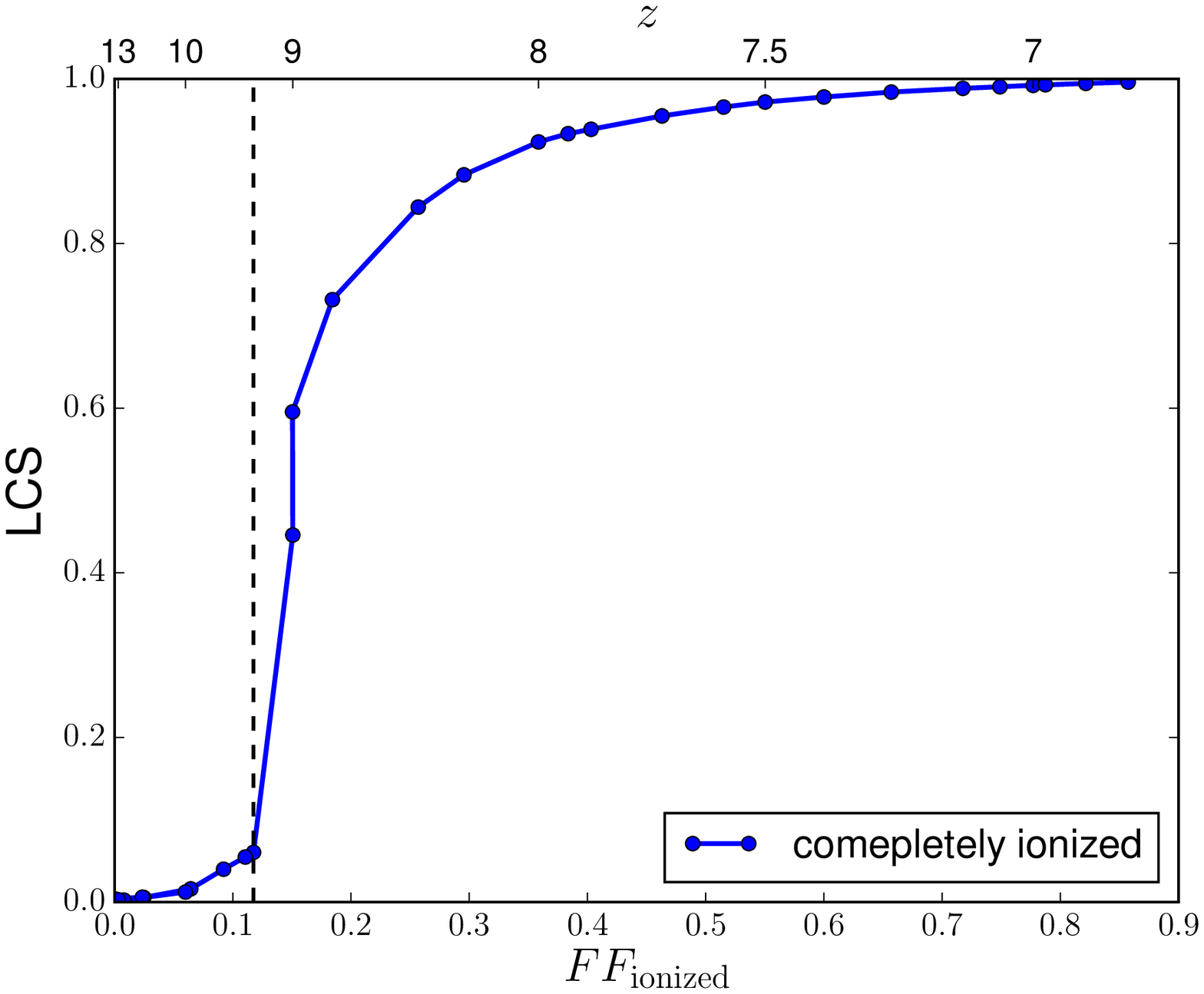}\label{fig:lcs_vs_ffc}}
\caption{{\bf (a):} Largest Cluster Statistics (LCS) of the ionized regions (with $\rhi=0$) and the 
neutral regions (with $\rhi>0$) are plotted against the neutral fraction $\xhi$. The corresponding redshift
is shown along the top x-axis. The largest neutral region is percolating in the entire range of study, $ 7\lesssim z \lesssim 13$. On the other hand, for the ionized regions, the percolation transition takes place at $\xhi \approx 0.728$, 
which roughly corresponds to $z \approx 9$, and is shown by the vertical black dashed line. During the percolation transition the LCS  undergoes a sharp rise, as shown in the left panel. {\bf (b):} The LCS for the ionized regions is plotted against the (ionized) filling factor, $\ffc$. The largest ionized region starts to percolate at $\ffc \approx 0.12$ and during the percolation transition LCS increases very steeply as well. We note that the filling factor, $\ffc$, is essentially the same as the volume weighted ionization
fraction.}
\label{fig:comp_ionized_lcs}
\end{figure*}
 
We have considered a number of HI density fields with neutral fraction ranging between $\xhi \in (0.1-1.0)$, where the lower limit, $\xhi=0.1$, corresponds to the redshift $z \approx 7$ and the upper limit $\xhi=1$ corresponds to high redshifts (before reionization was initiated). We study the fully ionized regions within the HI density field having $\rhi=0$. Side by side, we also consider the complementary region with
$\rhi>0$ and refer to it as the neutral segment. We define the individual regions, in both segments separately, as the connected grid points of same type (ionized or neutral) using the friends-of-friends (FoF) algorithm compatible with the periodic boundary condition, as explained in figure \ref{fig:pbc}. 

In percolation analysis, a key role is played by two quantities:
\begin{enumerate}

\item
The `largest cluster statistics' (LCS), defined for the ionized or the neutral segment as (\cite{Klypin1993})
\begin{equation}\label{eq:lcs}
  {\rm LCS}=\frac{\rm volume ~of ~the ~largest~neutral ~or ~ionized ~region}{\rm total ~volume ~of ~all ~the ~neutral ~or ~ionized ~regions}\;,
\end{equation}
is the fraction of the volume (ionized or neutral) filled by the largest region.
\item
The filling factor $FF$, which is defined for the ionized or the neutral segment as, 
\begin{equation}\label{eq:ffc}
 FF=\frac{\rm total ~volume ~of ~all ~the ~neutral ~or ~ionized ~regions}{\rm volume ~of ~the ~simulation ~box}\;.
\end{equation}
\end{enumerate}

 In the left panel of figure \ref{fig:comp_ionized_lcs} we plot the largest cluster statistics (LCS)  vs the neutral fraction ($\xhi$) for both the neutral (red) and ionized (blue) segments,
 while in right panel the plot of LCS vs the filling factor ($\ffc$)  is shown for the ionized segment only.

\begin{figure*}
\centering
\subfigure[$\xhi=0.793$, well before percolation]{
\includegraphics[width=0.323\textwidth]{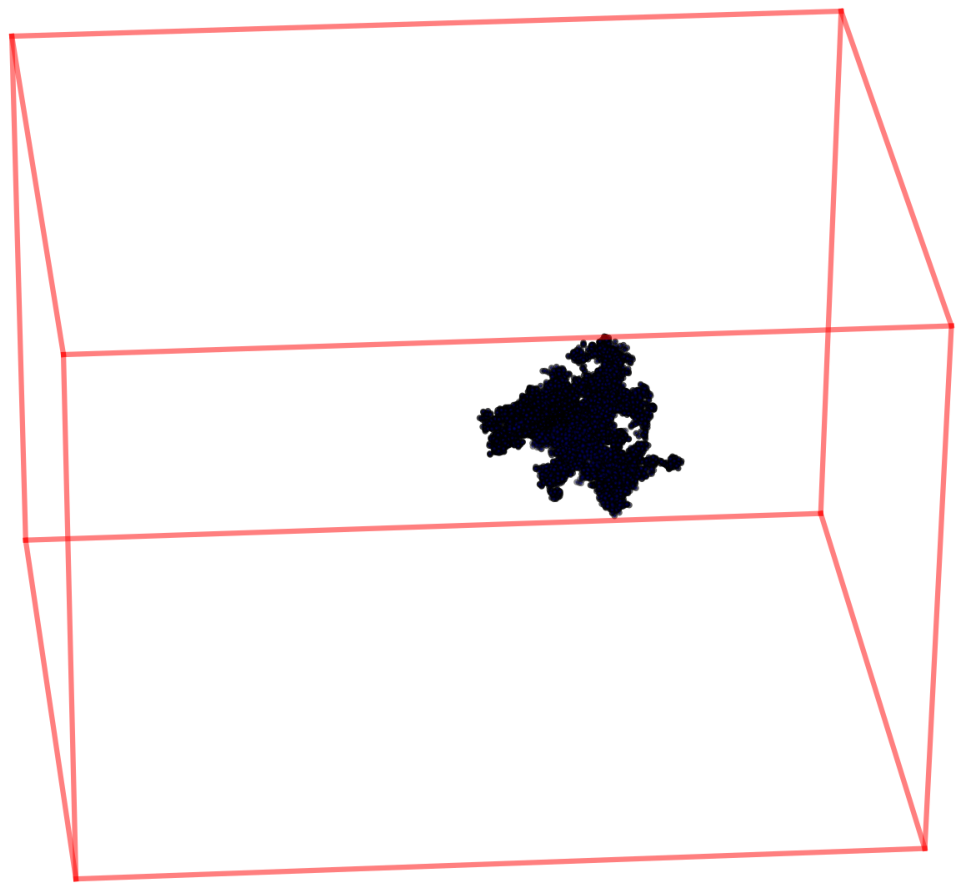}}
\subfigure[$\xhi=0.736$, before percolation]{
\includegraphics[width=0.323\textwidth]{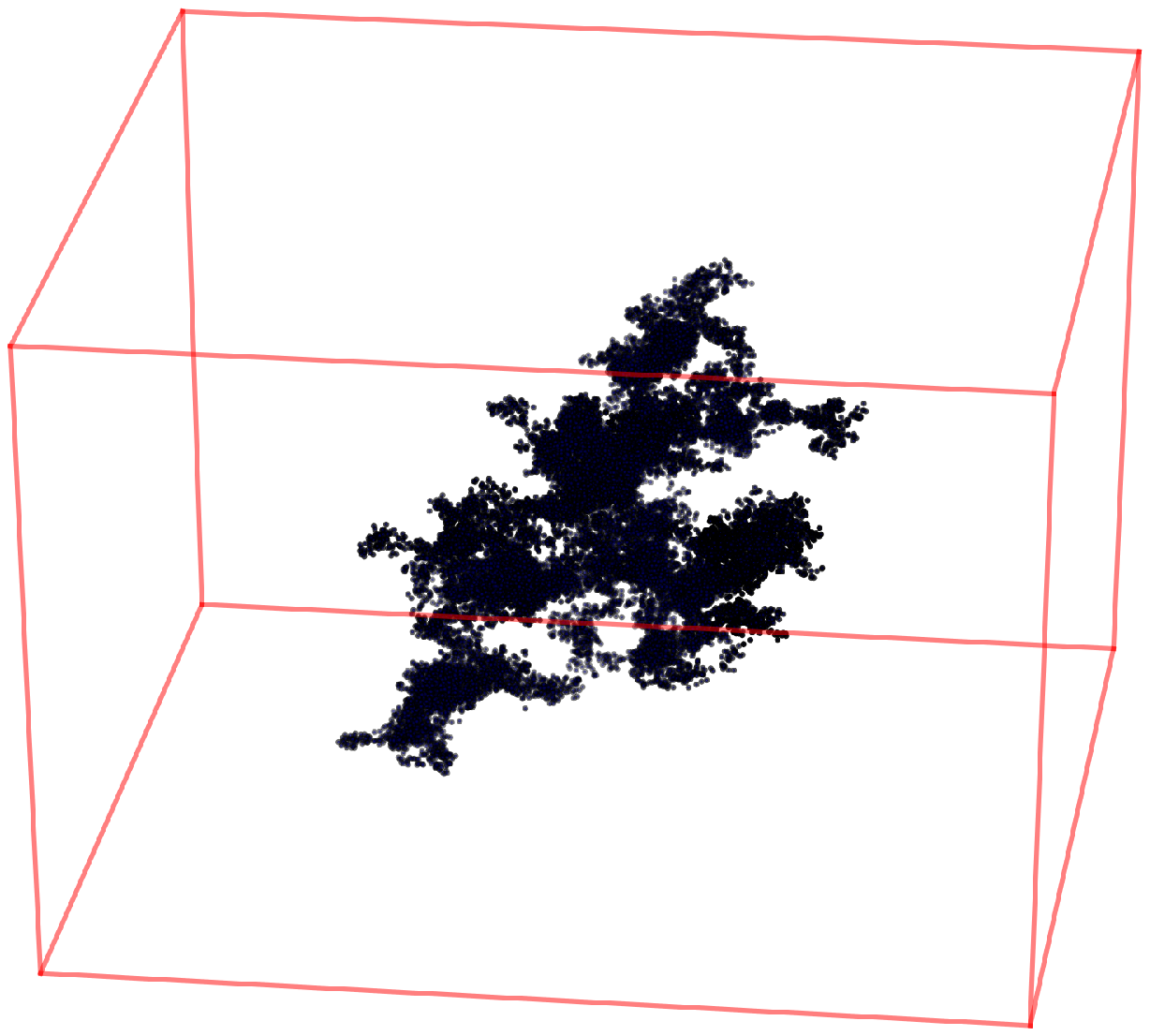}}
\subfigure[$\xhi=0.726$, just after percolation ($z=9$)]{
\includegraphics[width=0.323\textwidth]{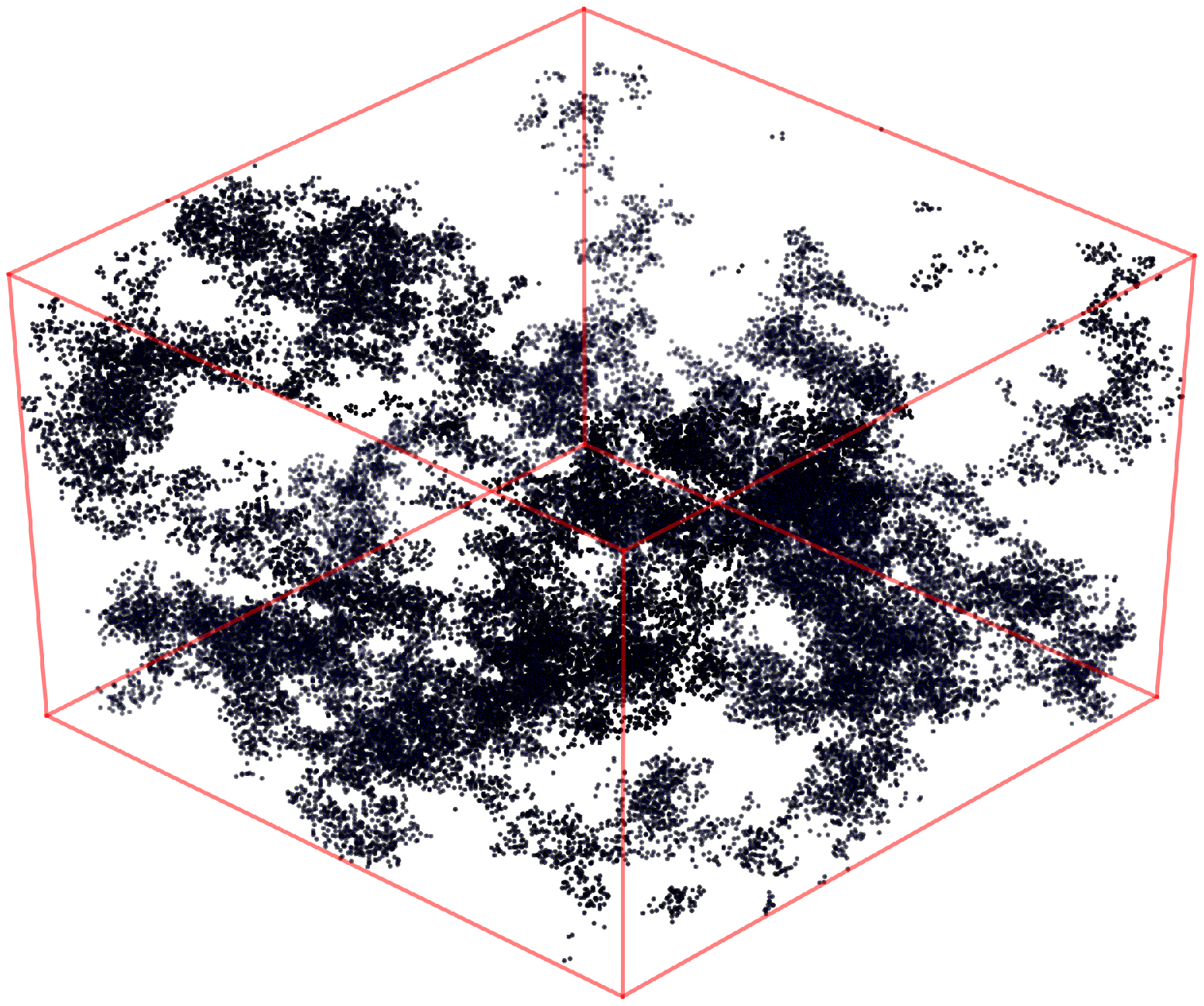}}
\caption{The largest ionized regions are shown within the simulation box (of dimension $\sim 215$ Mpc$^3$) for three values of the neutral fraction, $\xhi=0.793,~ 0.736,~ 0.726$ from the left. The corresponding ($\ffc$, LCS) are ($ 0.064$, $0.016$), ($ 0.117$, $0.061$) and ($ 0.151$, $0.446$) respectively. From the middle panel, it is visually evident that, just before percolation, the largest ionized region becomes like an inter connected filamentary structure with highly non-trivial topology. It becomes formally infinite in size beyond percolation.}
\label{fig:vis}
\end{figure*}

 As reionization progresses, the ionized segment grows in size and so does the largest ionized region. 
Soon the largest ionized region becomes so large that it stretches from one face of the simulation 
box to the opposite face. (Note that due to periodic boundary conditions such a region is formally
 infinite in size.) We refer to this as the `percolation transition'. During the percolation transition, the LCS increases sharply when plotted against $\xhi$ or $\ffc$, as shown in figures \ref{fig:lcs_vs_xhi} and 
\ref{fig:lcs_vs_ffc}. Indeed, the percolation transition itself can be identified through this abrupt rise in LCS (\cite{Klypin1993}). From both panels of figure  \ref{fig:comp_ionized_lcs} one finds that  
the ionized regions percolate for $z \lesssim 9$ (or equivalently $\xhi \lesssim 0.728$, $\ffc \gtrsim 0.12$). These critical thresholds at percolation appear to be quite stable for simulations with different resolutions \footnote{It would be interesting to study how the percolation transition statistically changes for different models of reionization. Since enormous computational time would be required to carry out this comparison, we leave the exercise for a follow up project.}. After percolation, most of the individual ionized regions are rapidly assimilated into the largest ionized region, for example at $\xhi \approx 0.55$ or $\ffc \approx 0.3$, almost $90 \%$ of the ionized hydrogen resides in the largest ionized region. Therefore, at percolation there is a sharp transition from individual ionized regions to one large connected ionized region. The latter forms by the overlap or merger of the individual ionized regions.
This is also evident from figure \ref{fig:vis} where the largest ionized regions are visualized within the simulation box (of dimension $\sim 215$ Mpc$^3$) at three different values of the neutral fraction corresponding to well before, just before and just after percolation (from left).
These results are completely 
consistent with earlier works, \cite{Iliev2006,Chardin2012,Furlanetto2016}, where different simulation methods were used to generate the HI density fields.   

We also find that the neutral segment is percolating (LCS being close to unity) during the entire redshift range under study, namely
$ 7 \lesssim z \lesssim 13$. Therefore, in the range $7 \lesssim z \lesssim 9$ (or equivalently $0.1 \lesssim \xhi \lesssim 0.728$), both the neutral and ionized parts of the hydrogen field are
infinitely extended through their connected regions. As we know, after reionization, the neutral hydrogen remains confined to galaxies and local clusters suggesting that the percolation transition in HI takes place within the range $6<z<7$. In the rest of the paper we focus only on the ionized part, namely HII.

If the ionized bubble size obeys a pure power law distribution,  
\begin{equation}\label{eq:n_dist}
 \left(\frac{dN}{dV'}\right) \propto V^{\tau}\;,
\end{equation}
then
\begin{equation}\label{eq:NV}
 \mathcal{N}(V) \equiv \int_V ^{\infty} \left(\frac{dN}{dV'}\right) dV'=C V^{1+\tau}\;.
\end{equation}
 is the (cumulative) number of distinct ionized regions, each having minimum volume $V$. The fraction of ionized volume filled by these $\mathcal{N}(V)$ regions is given by,
\begin{equation}\label{eq:fV}
 \mathcal{F}(V) \equiv \int_V ^{\infty} V'\left(\frac{dN}{dV'}\right) dV' \approx \frac{ \left (V_{max}^{2+\tau}-V^{2+\tau} \right)}{\left( V_{max}^{2+\tau}-1 \right)}~~{\rm for}~~V<V_{max}\;.
\end{equation}

\begin{figure*}
\centering
\subfigure[]{
\includegraphics[width=0.47\textwidth]{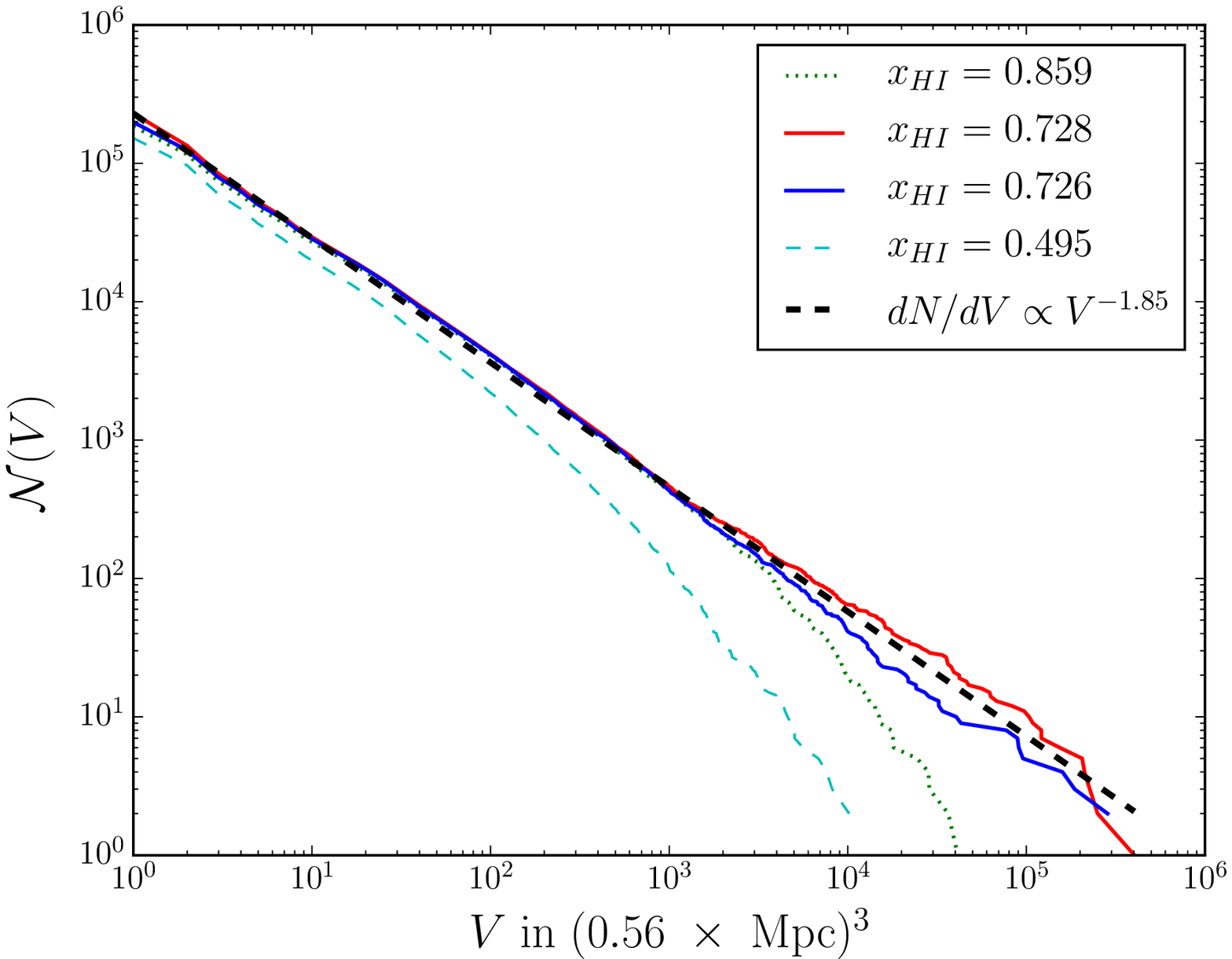}\label{fig:NV_V_fit}}
\subfigure[]{
\includegraphics[width=0.47\textwidth]{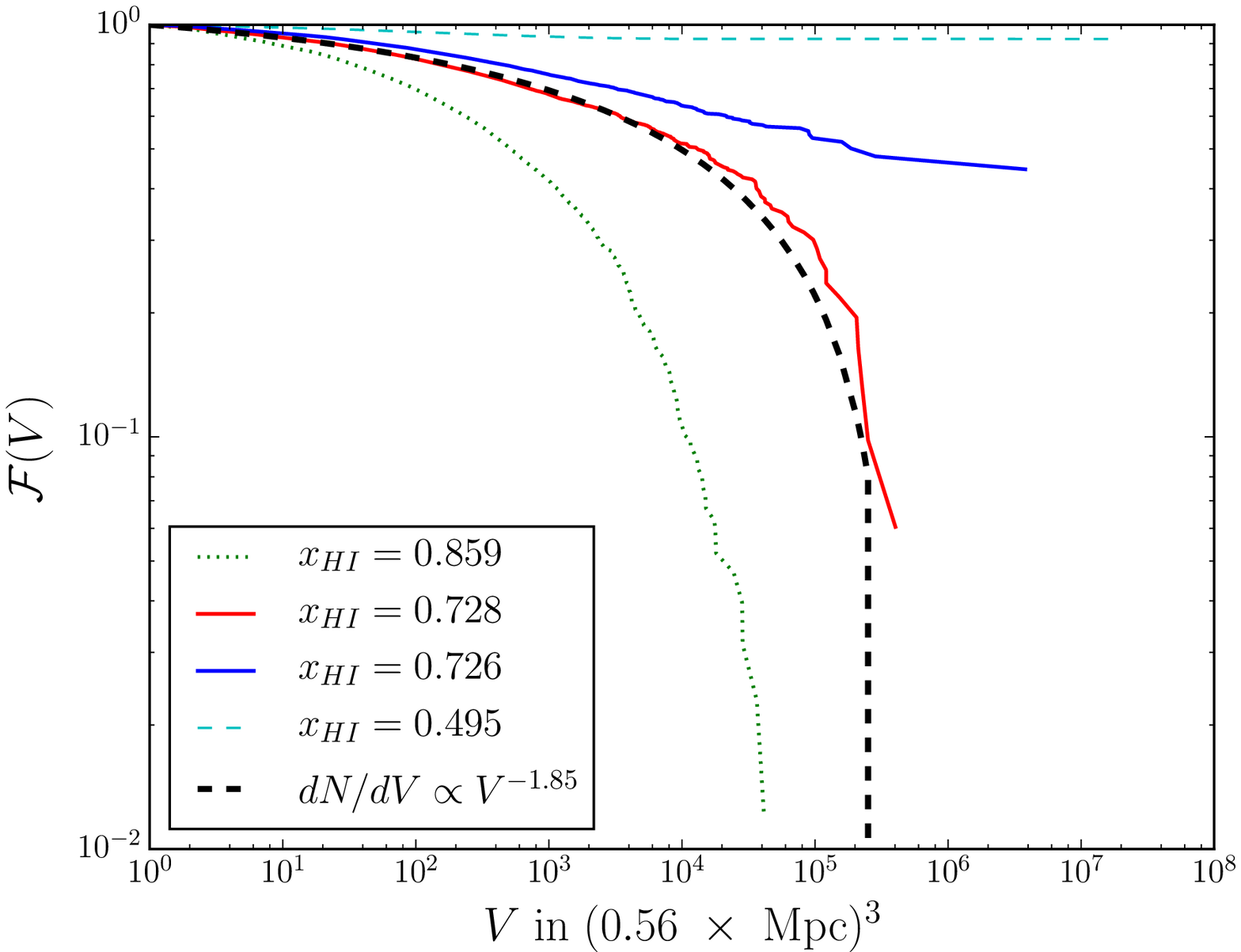}\label{fig:fV_V_fit}}
\caption{Bubble size distribution: {\bf (a):} The cumulative no. of region $\mathcal{N}(V)$, defined in \eqref{eq:NV}, is plotted against ionized region volume for different values of the neutral fraction. The solid curves show the distribution just before (red) and just after (blue) percolation while the green dotted and cyan dashed curves correspond to well before and well after percolation transition respectively. In this plot, we exclude the largest ionized region after percolation transition. The curves follow the power law distribution, $\mathcal{N}(V) \propto V^{1+\tau}$ with $\tau \approx -1.85$ (shown by the black dashed line), specially in the vicinity of percolation transition. The distribution with $\tau \approx -1.85$ is slightly less steep than what \protect\cite{Furlanetto2016} found, $\tau \approx -2$. {\bf (b):} The fraction of ionized volume filled by the regions with minimum volume $V$, is plotted for the same set of neutral fractions. We observe that, well after percolation, most of the ionized volume is filled by the largest ionized region. For example, the cyan dashed line shows that at $\xhi=0.495$ (corresponds to $z=8$), the percolating region encloses almost $92 \%$ of the ionized volume. On the other hand, well before percolation, most of the ionized volume is distributed in smaller regions. The black dashed line corresponds to the power law distribution, given in \eqref{eq:n_dist} with $\tau \approx -1.85$, which matches well with the curve corresponding to just before percolation. Note that here we estimate the volume of each ionized region by counting the grid points inside it. Therefore, the smallest ionized regions have only one grid point inside them which roughly corresponds to volume $V \sim (0.56$ Mpc$)^3$.}
\label{fig:bubble_dist}
\end{figure*}

In figure \ref{fig:bubble_dist}, we plot the bubble size distribution \footnote{In section \ref{sec:cir}, we calculate the volume of individual ionized regions by counting the grid points inside each region. The number of grid points inside an ionized region roughly reflects its volume. One could have used SURFGEN2, explained in section \ref{sec:shape}, to calculate the volume of each individual region more precisely, but that would increase the computation time enormously. Moreover, for extremely small ionized regions SURFGEN2 won't be very precise because of finite grid effects.}. The solid curves are corresponding to just before percolation (red) and just after percolation (blue) in both the panels while the green dotted and cyan dashed curves are for well before and well after percolation respectively. $\mathcal{N}(V)$, shown in figure \ref{fig:NV_V_fit}, roughly obeys the power law distribution, given in \eqref{eq:NV} with $\tau \approx -1.85$, over a large volume range. Clearly the power law distribution is more accurate near the percolation transition. These results are overall in accordance with \cite{Furlanetto2016}, but the slope ($\tau \approx -1.85$) in our finding is slightly less compared to their result, $\tau \approx -2$ . The power law distribution ensures that most ionized regions are very small in size without any existence of characteristic bubble size, which has been investigated extensively in literature (\cite{Iliev2006,Friedrich2011,Lin2016, Kakiichi2017}). Figure \ref{fig:fV_V_fit} shows behaviour of $\mathcal{F}(V)$, the fraction of ionized volume filled by regions of volume $V$ and higher (defined in \eqref{eq:fV}), with the volume $V$. It is clearly evident that at lower $\xhi$ beyond percolation, most of the completely ionized volume is enclosed by the largest ionized region. On the other hand, at higher $\xhi$, smaller regions fill most of the ionized volume. Again the power law distribution with $\tau \approx -1.85$ matches quite well with the $\mathcal{F}(V)$ at the onset of percolation.

\section{Determining the shapes of ionized regions using {\em Shapefinders}}\label{sec:shape}

\begin{figure}
\centering
\includegraphics[width=0.47\textwidth]{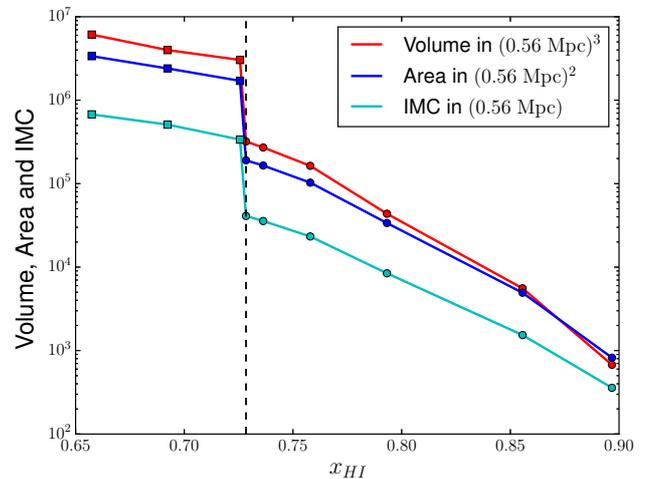}
\caption{ The Minkowski functionals, namely volume, area and integrated mean curvature (IMC), of the largest ionized region are plotted against neutral fraction ($\xhi$) in the vicinity of percolation transition. The percolation transition is shown by the dashed vertical line and the Minkowski functionals of the largest ionized region before and after percolation are shown by  filled circles and squares respectively. Since the largest ionized region grows rapidly during percolation, its volume , area and IMC increase sharply during percolation transition. All the three Minkowski functionals abruptly rise by almost an order of magnitude.  In this paper, all the values of Minkowski functionals and Shapefinders have been quoted in comoving scale.}
\label{fig:VAC_ionized}
\end{figure}

\begin{figure*}
\centering
\subfigure[]{
\includegraphics[width=0.47\textwidth]{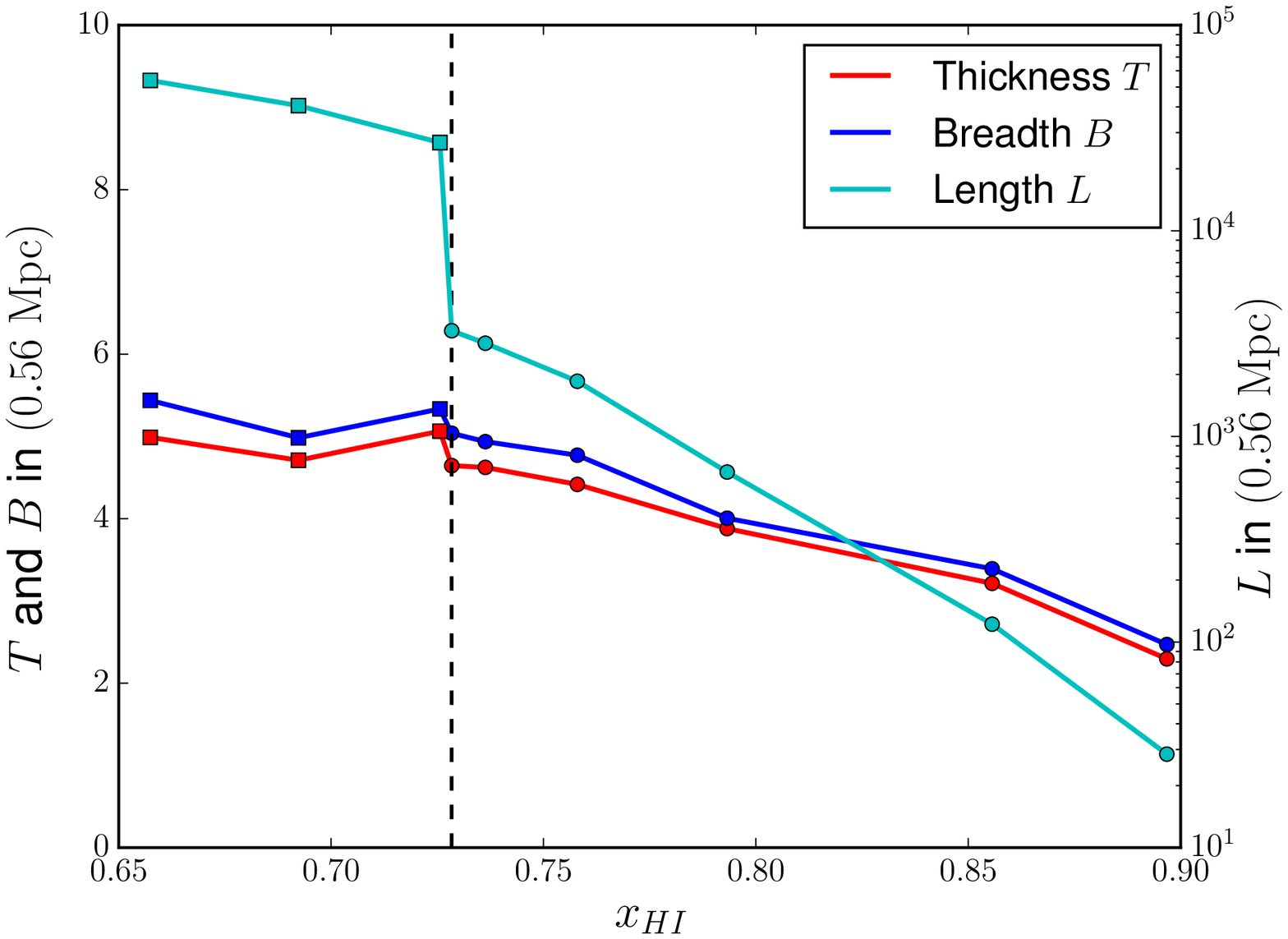}\label{fig:TBL_ionized}}
\subfigure[]{
\includegraphics[width=0.47\textwidth]{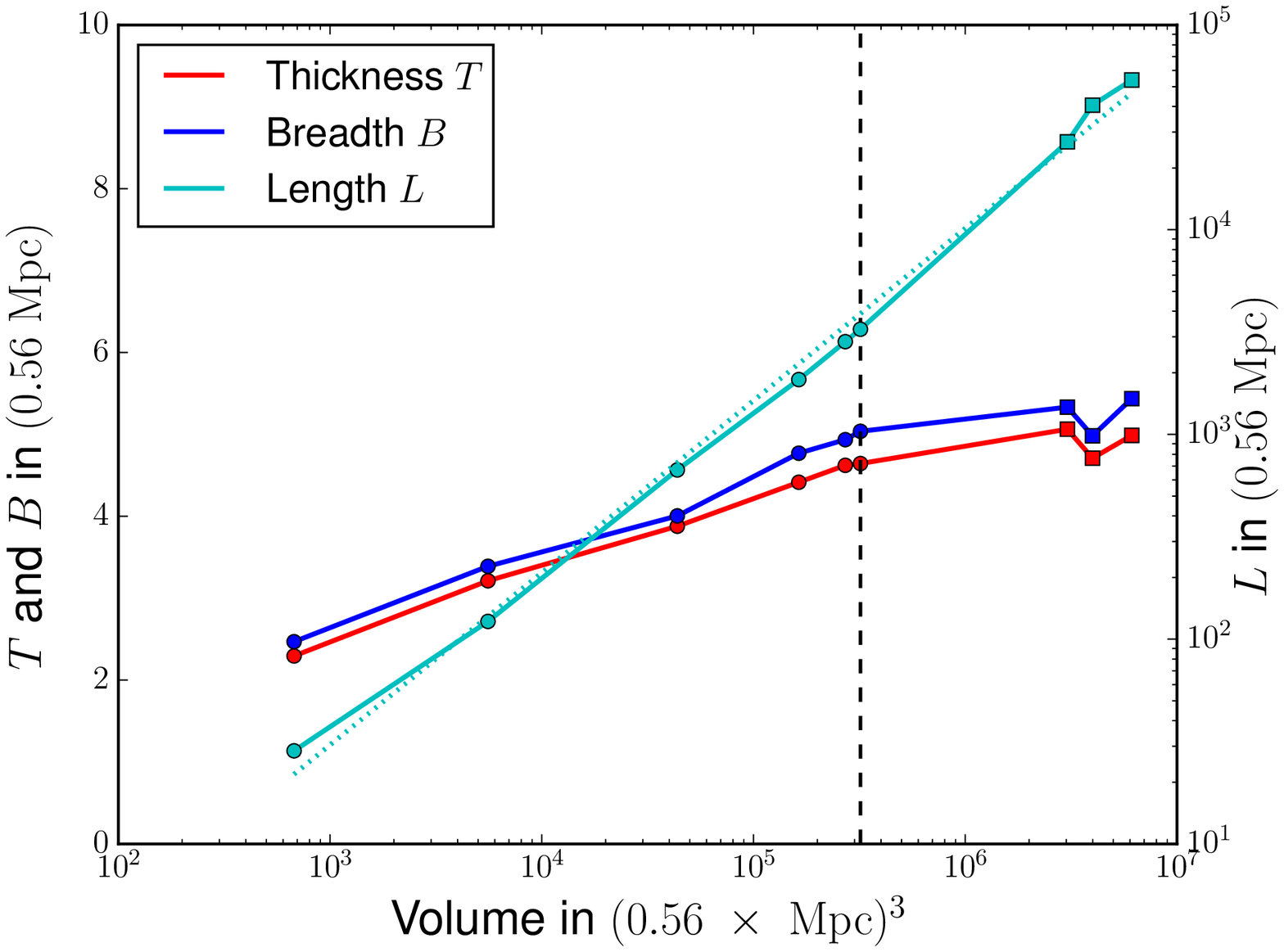}\label{fig:largest_TBL_vs_vol}}
\caption{{\bf (a)}  The volume, area and the integrated mean curvature of the largest ionized region grow in such a way that if one computes their ratios, the two Shapefinders -- thickness and  breadth -- do not rise much while the third Shapefinder -- length (plotted along right y-axis) -- increases rapidly near the percolation transition.  Note that, in this paper all the values of Shapefinders are shown in comoving scale. For the largest ionized region, one always obtains $T \approx B \ll L$ in the figures. Near the percolation transition, at $\xhi \approx 0.728$, $L \sim 10^3 B$. {\bf (b)} The Shapefinders of the largest ionized region are plotted against the region's volume as the latter grows during reionization near percolation. The percolation transition is shown by the vertical dashed line in both the panels. The thickness and breadth increase very slowly with volume while length increases almost as power law. The slope of the best fit straight line to $\log L$ vs $\log V$ curve, shown by the dotted cyan line, is very close to unity; $m_L=0.841$. Hence the length of the largest ionized region increases almost linearly with volume in the vicinity of percolation while the cross-section, estimated by $T \times B$, does not vary much. Therefore, both the panels show that the largest ionized region possesses a characteristic cross-section of $\sim 7$ Mpc$^2$ during its rapid growth near the percolation. Note the enormous difference between $T$, $B$ on the one hand and $L$ on the other, there is no real intersection of $L$, $B$, $T$, as apparently suggested by the figures.}
\label{fig:largest_TBL}
\end{figure*}

\begin{figure}
\centering
\includegraphics[width=0.47\textwidth]{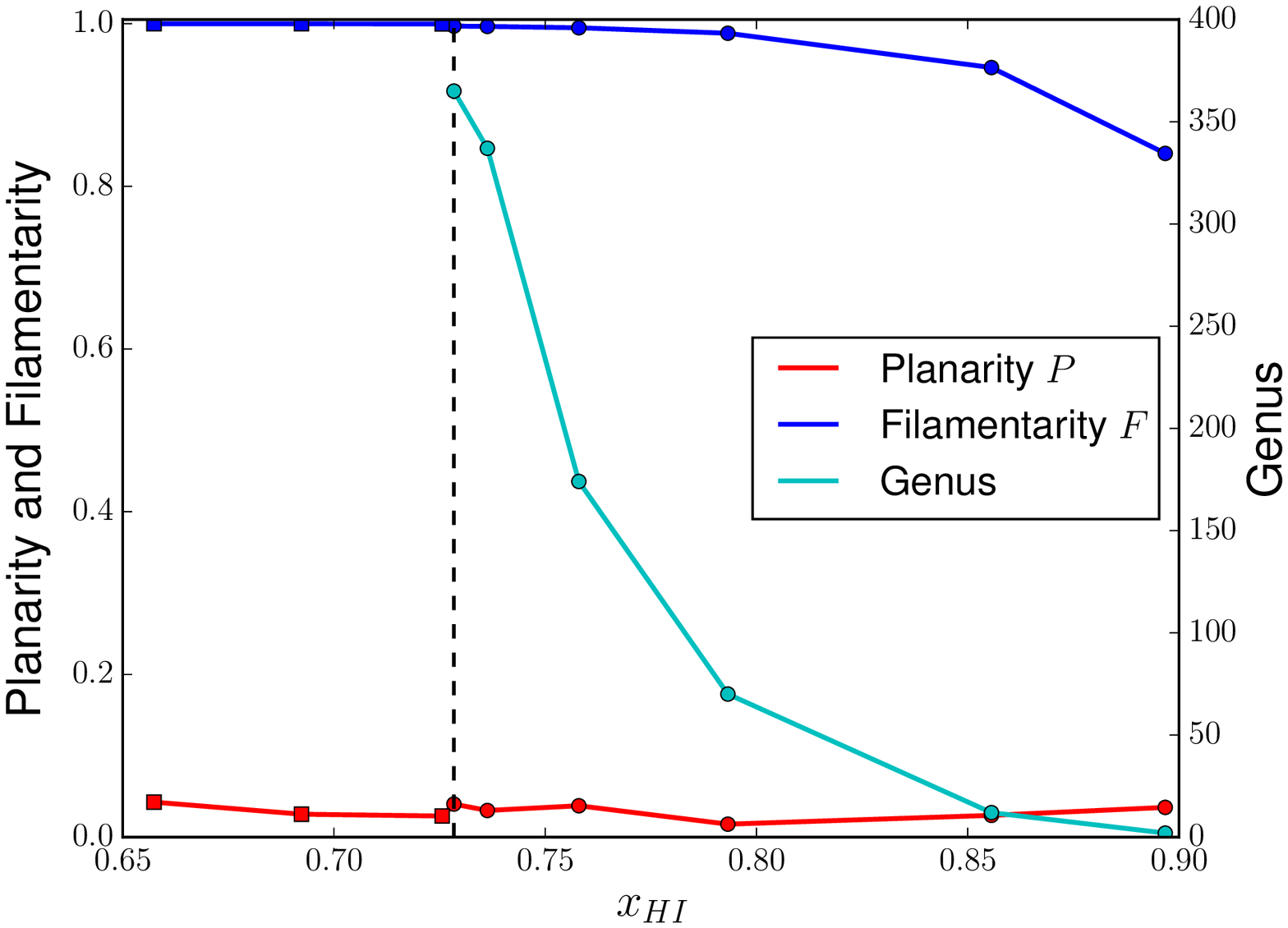}
\caption{The `planarity' ($P$), `filamentarity' ($F$) and genus of the largest ionized region are shown as a function of the neutral fraction ($\xhi$) in the vicinity of the percolation transition. The filamentarity of the largest ionized region rises to almost unity during percolation but the planarity remains quite small. The genus of the largest ionized region also increases as the region grows. 
Therefore, the largest ionized region becomes very filamentary and more tunnels pass through the largest ionized region as the latter grows rapidly at the onset of percolation. 
Note that we do not show the genus value of the largest ionized region after percolation ($\xhi < 0.728$), due to uncontrollable errors which arise because of periodic boundary condition.
}
\label{fig:PF_comp_ionized}
\end{figure}

In this section we study the shapes of the ionized regions at different redshifts (various stages of reionization) using Shapefinders, which are derived from Minkowski functionals.  
The morphology of a closed two dimensional  surface embedded in three dimensions 
is well described by the four Minkowski functionals (\cite{mecke})
\begin{itemize}
 \item Volume: $V$,
 
 \item Surface area: $S$,
 
 \item Integrated mean curvature (IMC):
 \begin{equation}
  C=\frac{1}{2} \oint \left(\kappa_1+\kappa_2\right) dS \;,
 \end{equation}
 
  \item Integrated Gaussian curvature or Euler characteristic: 
  \begin{equation}
  \chi=\frac{1}{2\pi} \oint (\kappa_1 \kappa_2) dS\;.
  \end{equation}

\end{itemize}
Here $\kappa_1$ and $\kappa_2$ are the two principle curvatures at any point on the surface. 
The fourth Minkowski functional (Euler characteristic) can be written in terms of the genus (G) 
of the surface as follows,
\begin{equation}
 G=1-\chi/2 \equiv {\rm (no.~ of~ tunnels)}-{\rm (no.~ of~ isolated~ surfaces)}+1\;.
\end{equation}
It is well known that $\chi$ (equivalently $G$) is a  measure of the topology of the surface.

The `Shapefinders', introduced in \cite{Sahni:1998cr}, are ratios of these Minkowski functionals,
namely
\begin{itemize}
 \item Thickness: $T=3V/S$,
 \item Breadth: $B=S/C$\;,
 \item Length: $L=C/(4\pi)$. 
\end{itemize}

The Shapefinders $T, B, L$, have dimension of length, and can be interpreted as 
providing a measure of the three physical dimensions of an object
\footnote{In general one finds $C >0$. However in the rare case when a region has 
$C < 0$ we shall redefine $C \to |C|$ to ensure that $B$ and $L$ are positive. 
Furthermore, if the natural order $T\leqslant B \leqslant L$ is not maintained, we choose the smallest dimension as $T$ and the largest one as $L$.}.
The Shapefinders are spherically normalized, i.e. $V=(4\pi/3) TBL $. 

Using the Shapefinders one can determine the morphology of an object (such as
an ionized region),
by means of the following dimensionless quantities 
 \footnote{One can redefine `Length' by taking the genus ($G$) of an object into account (\cite{Sheth:2002rf}),
\begin{equation}\label{eq:L1}
L_1=\frac{C}{4 \pi (1+|G|)}\;.
\end{equation}
 This reduces the filamentarity in the following manner while keeping planarity unchanged,
 \begin{equation}\label{eq:f1}
  F_1=\frac{L_1-B}{L_1+B}\;.
 \end{equation}}
 which characterize
 its planarity and filamentarity (\cite{Sahni:1998cr})
\begin{equation}\label{eq:PF}
 {\rm Planarity:}~ P=\frac{B-T}{B+T}\;, ~~\\
{\rm  Filamentarity:}~ F=\frac{L-B}{L+B}.
\end{equation}
For a planar object (such as a sheet) $P \gg F$, while the reverse is true for a filament which has
$F \gg P$. A ribbon will have $P \sim F \gg 0$ whereas $P \simeq F \simeq 0$ for a sphere.
In all cases $ 0 \leq P,F \leq 1$.

To calculate the Minkowski functionals and the Shapefinders of individual ionized regions, we developed a highly sophisticated code, named SURFGEN2, which models the surfaces of an ionized region through triangulation using {\em Marching Cube 33} algorithm (\cite{marcube,mar33}). SURFGEN2 is an advanced version of the SURFGEN algorithm, developed by \cite{Sheth:2002rf}. The detailed algorithms are described in \cite{Sheth:2002rf,bag2018b}. 
The accuracy of SURFGEN2 is excellent and much better than the existing methods of estimating the Minkowski functionals (\cite{Schmalzing 1997}), for example using the Koenderink invariant (\cite{Koenderink}) or the Crofton’s formula (\cite{Crofton_1968}).

\subsection{Shape of the largest ionized region during the percolation transition}

Figure \ref{fig:VAC_ionized} shows how the Minkowski functionals of the largest ionized region evolve during percolation transition, at around $\xhi \approx 0.728$. During the percolation transition, as the largest ionized region suddenly grows bigger, its Minkowski functionals, namely volume, area and integrated mean curvature (IMC), increase sharply\footnote{In this paper, all the values of Minkowski functionals and Shapefinders have been quoted in comoving scale.}. But the largest ionized region evolves in such a manner that two of its Shapefinders (ratios of Minkowski functionals), `thickness' $T$ and `breadth' $B$,
increase slowly as reionization proceeds. In fact these two quantities remain almost constant across the percolation transition. 
In contrast, the third Shapefinder, `length' $L$, increases steeply as reionization proceeds (see figure \ref{fig:TBL_ionized}) and increases  sharply by nearly an order of magnitude at the percolation transition. During the percolation transition, the largest ionized region abruptly grows only in `length' while the `cross-section', estimated by $T \times B$, does not change much. One might note that $L \sim 10^3 B$ near the percolation transition. From \eqref{eq:PF}, this implies that the largest ionized region is highly filamentary, with `filamentarity' $F \sim \mathcal{O}(1)$, near percolation. In figure \ref{fig:largest_TBL_vs_vol}, the Shapefinders of the largest ionized region are plotted against the region's volume as the latter grows in the vicinity of the percolation transition. Since $V \propto (T \times B \times L$), $L$ increases almost linearly with volume near percolation  where $T$ and $B$ increase much more slowly. The slope of the best fit straight line to the $\log L$ vs $\log V$ curve, shown by the dotted cyan line in figure \ref{fig:largest_TBL_vs_vol}, is of order unity; namely $m_L \equiv \log L/\log V \approx 0.841$. For comparison,  $m_L \approx 1/3$ for spherical surfaces, $m_L \approx 1/2$ for sheets and $m_L \approx 1$ for filaments. Hence we conclude that the largest ionized region possesses a characteristic cross-section ($\sim 7$ Mpc$^2$) which remains  almost 
constant across the percolation transition, 
while its length  grows rapidly near percolation.   

As illustrated in figure \ref{fig:PF_comp_ionized}, the `filamentarity' $F$ of the largest ionized region increases reaching almost unity near the percolation transition while the `planarity' $P$ is quite low and does not vary much. 
Therefore, the largest ionized region start to become highly filamentary at the onset of percolation. 
The genus of the largest ionized region, plotted along the right y-axis in figure \ref{fig:PF_comp_ionized},  also increases as reionization proceeds. This implies that as reionization proceeds the largest ionized
region acquires an increasingly complex topology  with many filamentary 
branches and sub-branches joining it, and several  tunnels passing through it. 


\subsection{Shapes of ionized regions during different stages of reionization}

\begin{figure*}
\centering
\subfigure[$\xhi=0.856$, well before percolation]{
\includegraphics[width=0.323\textwidth]{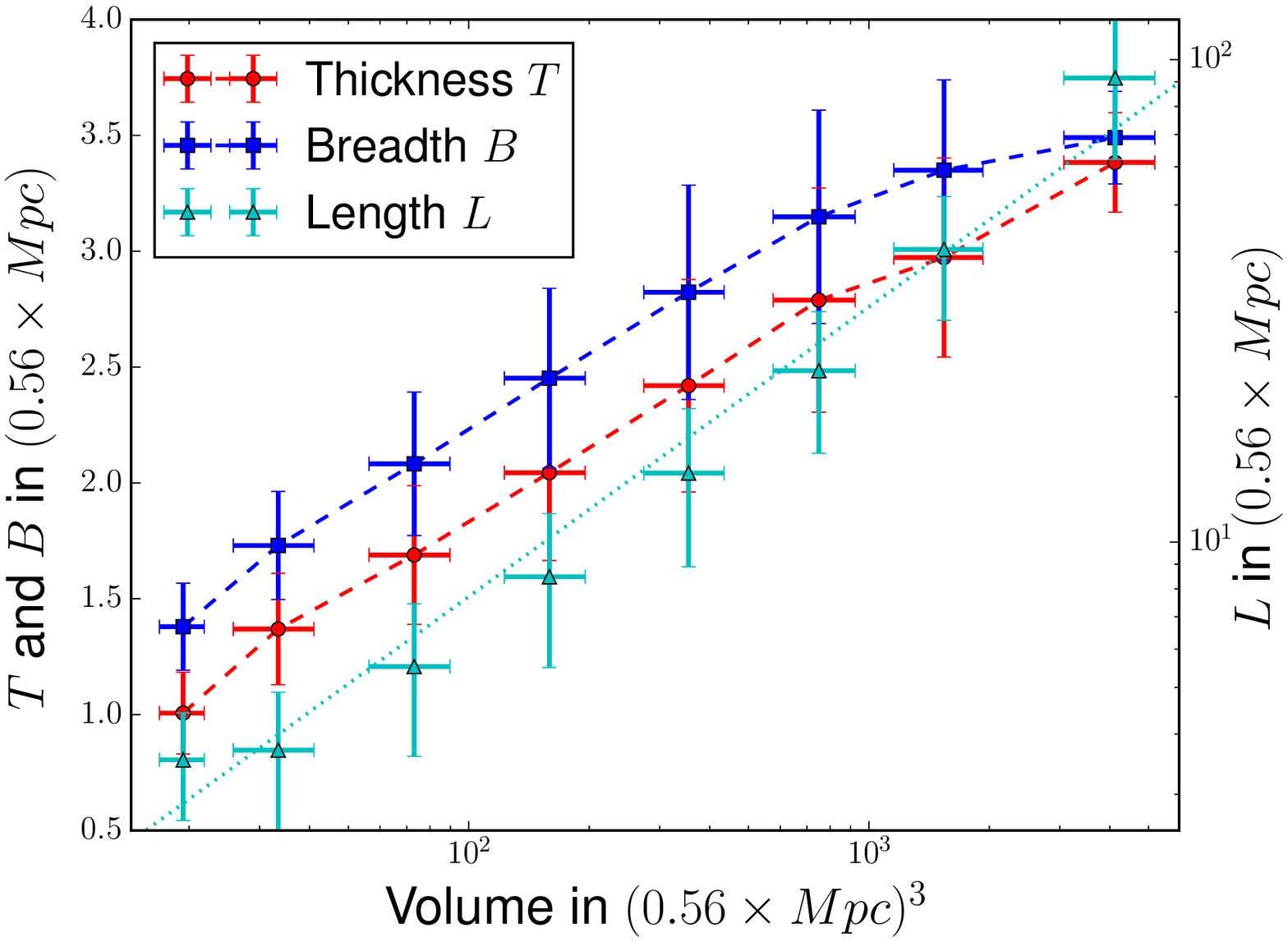}\label{fig:TBL_a}}
\subfigure[$\xhi=0.728$, just before percolation]{
\includegraphics[width=0.323\textwidth]{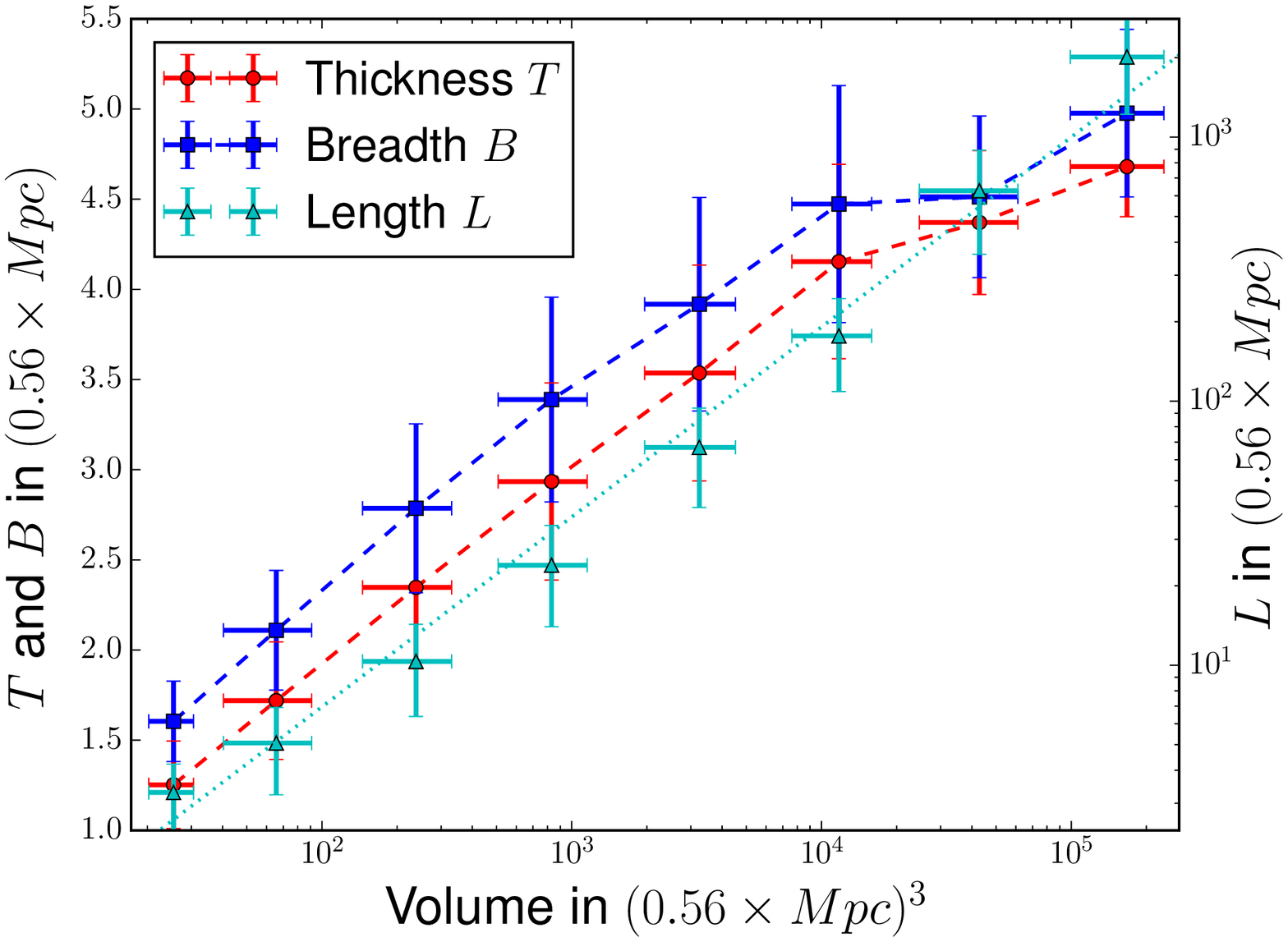}\label{fig:TBL_b}}
\subfigure[$\xhi=0.726$, just after percolation ($z=9$)]{
\includegraphics[width=0.323\textwidth]{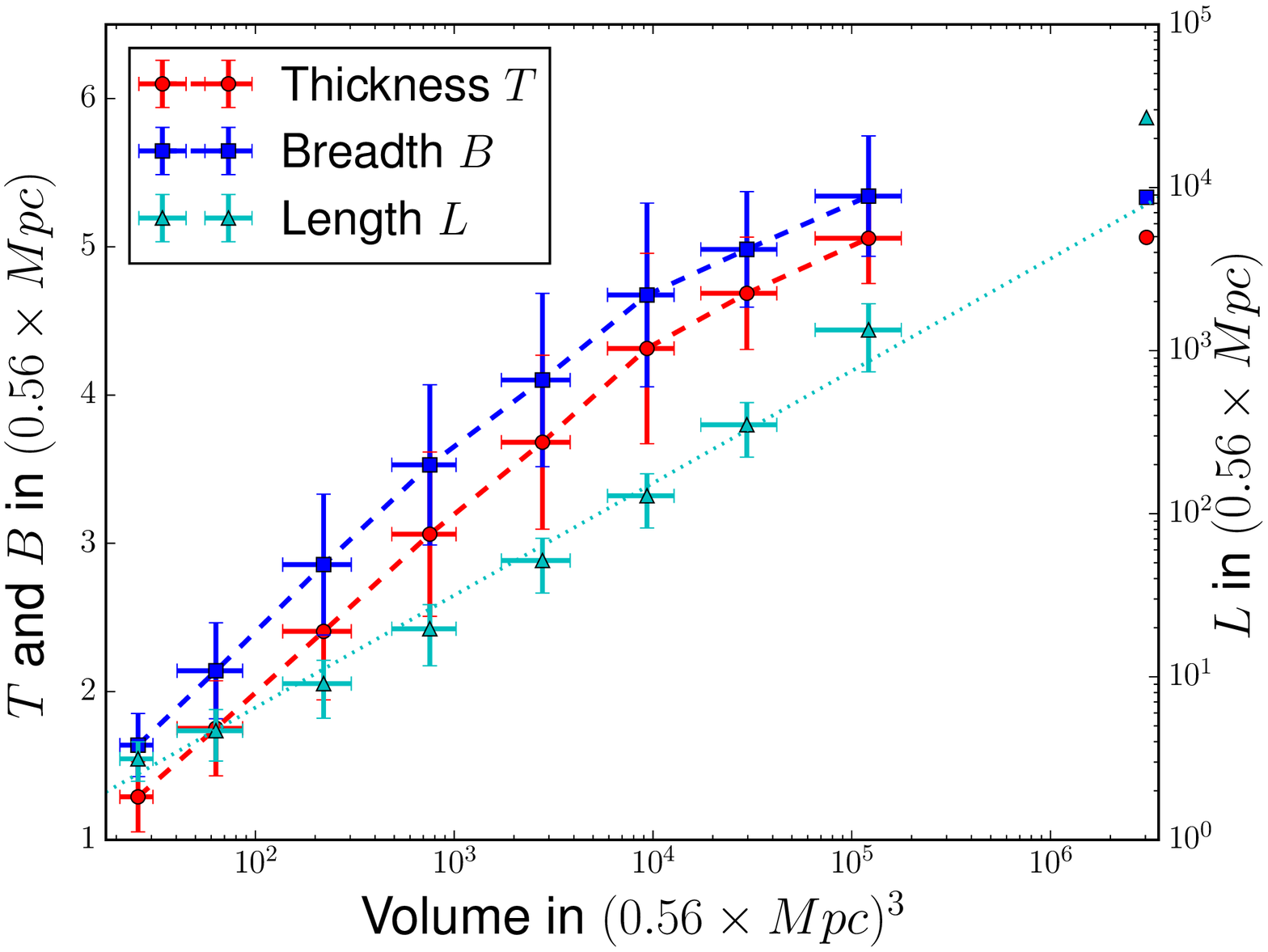}\label{fig:TBL_c}}
\caption{The Shapefinders `thickness' ($T$), `breadth' ($B$) and `length' ($L$) of ionized regions are plotted against their volume in bins, at three different values of neutral fraction. The error bars show standard deviations, which measure the scatter of ionized regions in each bin. The left panel corresponds to $\xhi=0.856$ which is well before percolation taking place in the ionized segment. Plots at the onset of percolation, for $\xhi=0.728$, are shown in the middle panel. The right panel shows the plots just after percolation, at $\xhi=0.726$ (which corresponds to $z=9$) where we have a large percolating ionized region. In all three figures, the thickness and breadth of ionized regions (plotted in linear scale along the left y-axis) increase much slowly with their volume when compared to the increase in their length with volume (plotted along right y-axis in log scale). We fit straight lines to $\log L$ vs $\log V$ curves, shown by the cyan dotted lines. The slope of the best fit straight line, $m_{L}$, increases as percolation is approached, for example $m_L \approx 0.60$ in the left panel for $\xhi=0.856$ while $m_L \approx 0.72$ at the onset of percolation, as shown in the middle panel. As the calculated Shapefinders of the percolating region in the right panel may suffer from accuracy, we exclude it from fitting. Note that different scales have been used in describing $T$, $B$ on the one hand and $L$ on the other. A very significant finding of our paper is that $T, B \ll L$ for large ionized regions in all three panels. This in turn implies that the large regions are filamentary, from \eqref{eq:PF}.}
\label{fig:TBL}
\end{figure*}

\begin{figure*}
\centering
\subfigure[$\xhi=0.856$, well before percolation]{
\includegraphics[width=0.323\textwidth]{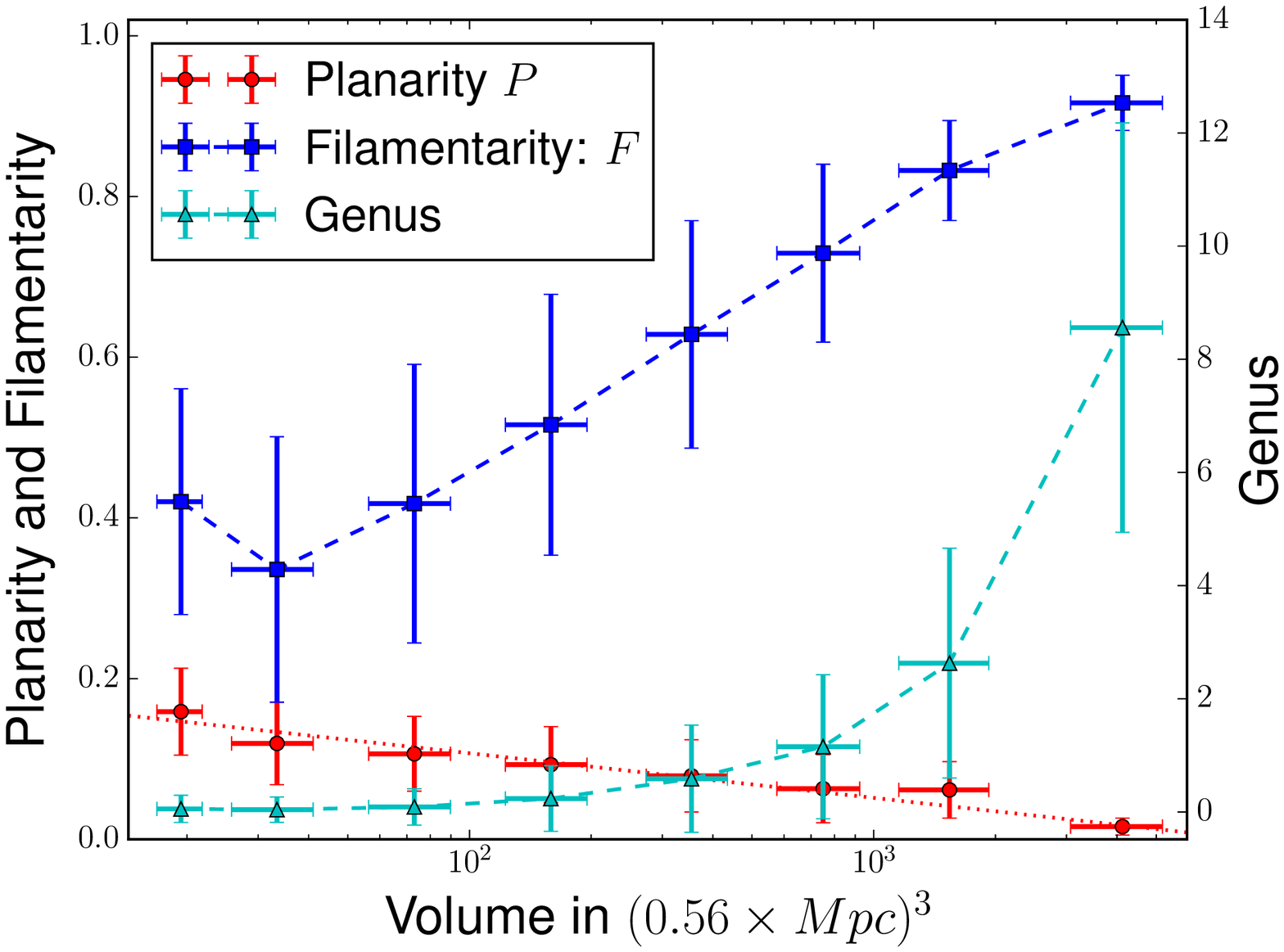}\label{fig:void_bin0_PF_z6_05_iso0_c50}}
\subfigure[$\xhi=0.728$, just before percolation]{
\includegraphics[width=0.323\textwidth]{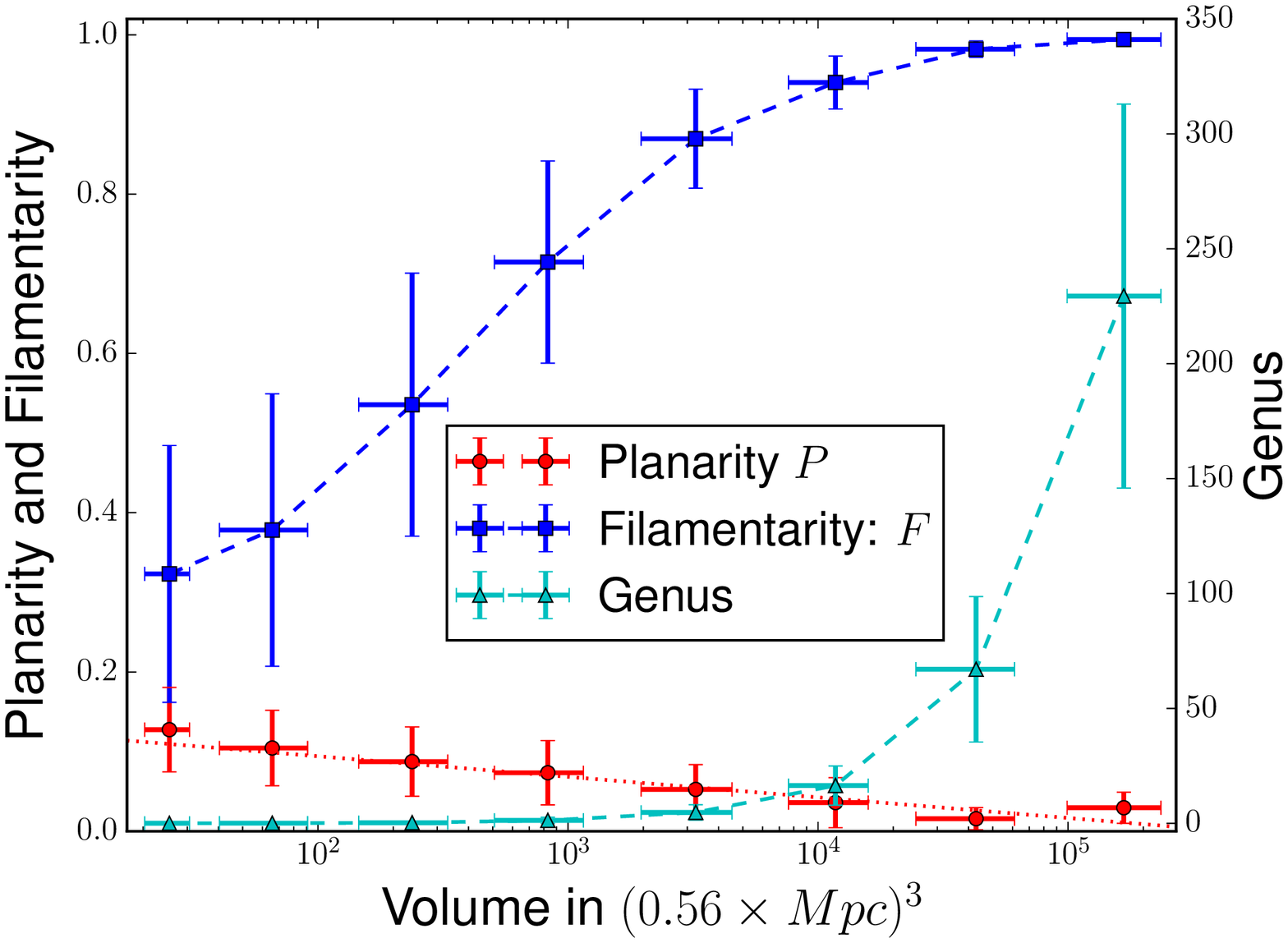}\label{fig:void_bin0_PF_z12_61_iso0_c50}}
\subfigure[$\xhi=0.726$, just after percolation ($z=9$)]{
\includegraphics[width=0.323\textwidth]{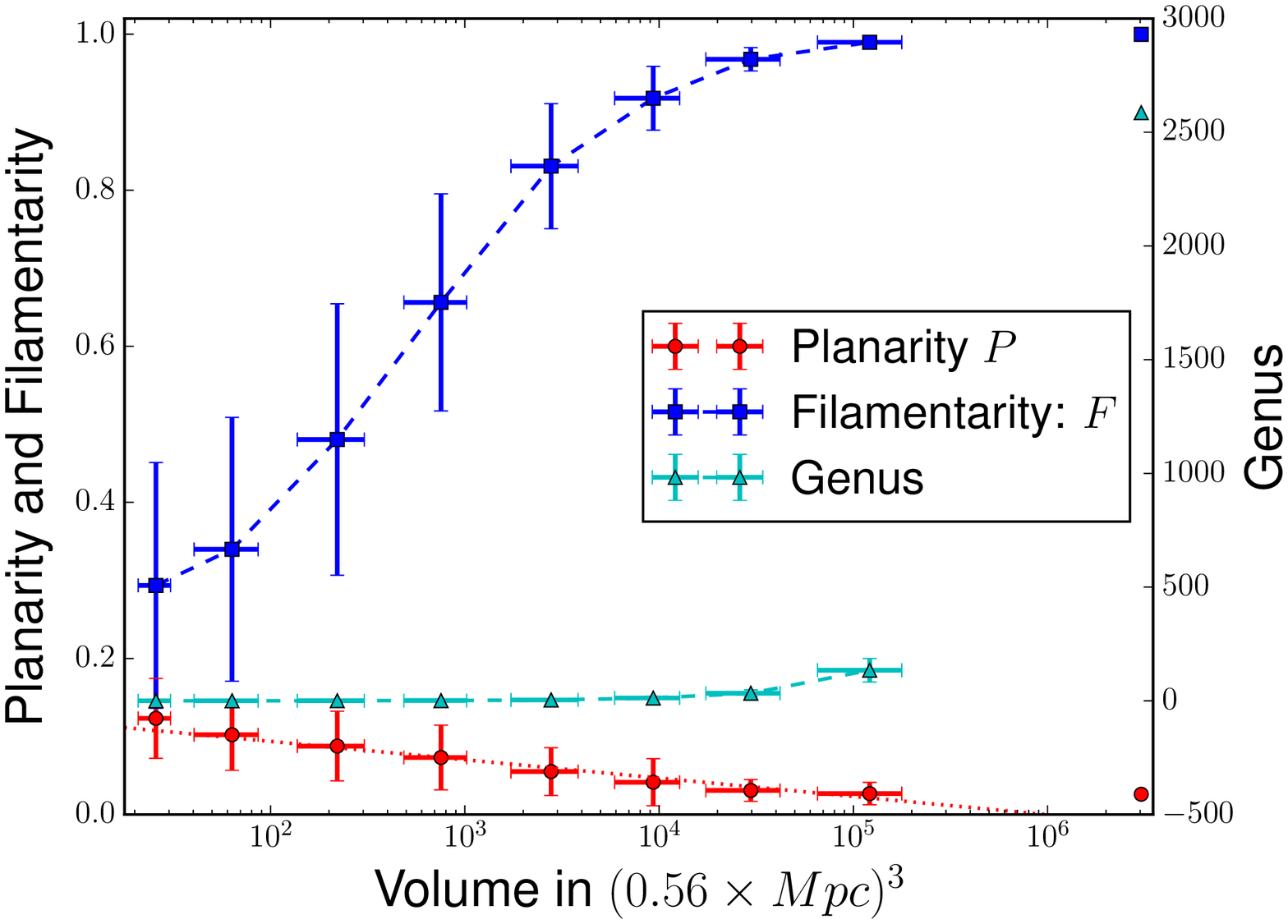}\label{fig:bin0_PF_z9_iso0_c50}}
\caption{To understand the shapes of the ionized regions, their `planarity' ($P$) and `filamentarity' ($F$) are plotted in volume bins at three different values of neutral fraction. The error bars show standard deviations, which measure the scatter of ionized regions in each bin. The left panel corresponds to $\xhi=0.856$ which is well before percolation taking place in the ionized segment. Plots at the onset of percolation, for $\xhi=0.728$, are shown in the middle panel.  The right panel shows the plots just after percolation, at $\xhi=0.726$ (which corresponds to $z=9$) where we have a large percolating ionized region. In all the three figures, the filamentarity of ionized regions increases with their volume. In contrast, the planarity of ionized regions does not change rapidly, in fact it decreases very slowly with increasing volume. Hence, the large ionized regions become very much filamentary. On the other hand, by extrapolating the curves to the lower volume, we observe that filamentarity and planarity (as well as genus) would be quite small which suggests that the most of the smaller regions have somewhat spherical morphology. But surprisingly, in the left panel (at very early stage of reionization), the filamentarity of extremely small ionized regions actually increases with decreasing volume. This indicates that at very early stages of reionization, most of the first small ionized bubbles are not exactly spherical but have trivial topology.}
\label{fig:PFG}
\end{figure*}

In this subsection we study the morphology of all ionized regions with Shapefinders at various redshifts. In principle one can calculate the Shapefinders of all the regions individually by triangulating their surfaces. However the surfaces of smaller regions, which have only a few grid points inside of them, cannot be accurately modeled by the triangulation scheme. Consequently their value of Minkowski functionals and Shapefinders suffer in accuracy due to low resolution. Moreover, since the number of ionized regions is very large near the percolation transition, calculating Shapefinders for all of them takes enormous computational time. Therefore, in this study we consider only sufficiently large ionized regions, with at least $50$ grid points inside each one, for the purpose of efficient triangulation. 

The Shapefinders, `thickness' ($T$), `breadth' ($B$) and `length' ($L$), of ionized regions are plotted against their volume in figure \ref{fig:TBL} for three different values of neutral fraction $\xhi$. The regions are binned in volume (equispaced bin width in log scale) and the error bars show standard deviations, which measure the scatter of ionized regions in each bin. The values of the neutral fraction $\xhi$ in the panels of figure \ref{fig:TBL} correspond to, commencing from the left: (a) well before percolation ($\xhi=0.856$), (b) just before percolation ($\xhi=0.728$) and (c) just after percolation ($\xhi=0.726$). In all cases one notices that the thickness ($T$) and the breadth ($B$) (plotted in linear scale along the left y-axis) of the large  ionized regions increase somewhat more slowly with volume when compared to the increase in length ($L$) with their volume (plotted along the right y-axis in log scale). This feature is more pronounced at the onset of percolation (middle panel). Since many large ionized regions appear as the percolation transition is approached, the cross-section of larger ionized  regions, measured by $(T \times B)$, are more alike near percolation.  We join the values of $T$ and $B$ in bins with dashed lines for visual guidance and fit  straight lines to $\log L$ vs $\log V$ curves. The slope of the best fit straight line (shown by the dotted line), $m_{L}$, increases as the neutral fraction approaches the critical value at percolation $\xhi^C \approx 0.728$. For example, $m_L \approx 0.60$ in the left panel for $\xhi=0.856$ while $m_L \approx 0.72$ at the onset of percolation for $\xhi=0.728$, as shown in the figures \ref{fig:TBL_a} and \ref{fig:TBL_b} respectively. Also the fit itself gets better near percolation; therefore large ionized regions show a power law dependence of length on their volume, which is more valid in the vicinity of percolation transition, just like the growth of the largest ionized region. After percolation, there exists a large percolating region , as shown in figure \ref{fig:TBL_c}. Since its Shapefinders may not be calculated accurately, we exclude it from fitting or joining. Well beyond percolation, the largest ionized region continues to grow and the rest of the ionized regions become smaller in number as well as in size. These scenarios are less important and have therefore been excluded from our Shapefinder analysis.

The planarity, filamentarity and genus of ionized regions are shown in figure \ref{fig:PFG}. The same set of values of neutral fraction is used as in figure \ref{fig:TBL}. The filamentarity and genus are joined by dashed lines for visual guidance while the planarity is fitted with a (dotted) straight line. It is interesting to note that the filamentarity of ionized regions increases with volume while the planarity slowly decreases with increasing volume. This explicitly demonstrate that large ionized regions are very filamentary around the percolation threshold.

The genus, plotted along right y-axis in figure \ref{fig:PFG}, also increases with volume, i.e. more tunnels pass through larger ionized regions. This suggests that large filamentary ionized regions are multiply connected with non-trivial topology.  The characteristic cross-section of these large regions is well described by   $T \times B \sim 7$ Mpc$^2$ near percolation.
One concludes that large ionized regions grow via the merging of relatively smaller ionized regions which were themselves large enough to be quite filamentary and to possess similar cross-sections. On the other hand, by extrapolating the curves towards lower volume, one finds that both filamentarity and planarity (as well as genus) of smaller regions can be quite low. Hence smaller regions are quite spherical with trivial topology.

In figures \ref{fig:TBL} and \ref{fig:PFG}, the standard deviations in each volume bin are shown by the respective error bars. It is evident from these figures that the error bars on the first two Shapefinders -- thickness ($T$) and breadth ($B$) -- as well as on the planarity ($P$) and filamentarity ($F$) shrink as we move to higher volume bins. As we know the large regions are formed by many interconnected filamentary branches and sub-branches (substructures) joining together. These substructures are large enough to possess similar values of $T$, as well as $B$. The Shapefinders, $T$ and $B$ of large ionized regions are somewhat averaged thickness and breadth of all these substructures. Therefore, $T$, as well as $B$, of larger ionized regions are more alike because the number of substructures is higher. In comparison, the smaller ionized regions have lesser number of substructures resulting in slightly more diverse $T$ and $B$. Hence, despite having fewer ionized regions in higher volume bin, the standard deviations (shown by the error bars) in $T$ and $B$, as well as in $P$, are smaller in higher volume bins. This leads to the characteristic cross-section in the large regions. On the other hand, the filamentarity ($F$) increases with region volume and $F$ of large regions are already very close to unity. Therefore, the scatter in $F$ is much less in higher volume bins. Smaller regions are lesser filamentary in general and have comparatively diverse morphology.
Note that the error bars on volume, length ($L$) and genus do not shrink in higher volume bins.

One intriguing fact to note is that in the early stages of reionization, the filamentarity of very small ionized regions actually increases slightly as we move to smaller ionized regions, see figure \ref{fig:void_bin0_PF_z6_05_iso0_c50}. This characteristic is quite robust and can be found in all HI density fields at early stages of reionization. Another interesting fact is that planarity of ionized regions slightly increases with decreasing volume at all stages of reionization. These features indicate that early ionized bubbles are not exactly spherical but mostly have trivial topology. Unfortunately, due to coarse resolution, we can not precisely calculate Shapefinders for extremely small ionized bubbles.


\section{Conclusion and discussions} \label{sec:conclusion}
Minkowski functionals and Shapefinders are powerful means of studying the geometry and topology 
 of large scale structures. We employ them in conjunction with percolation analysis, to study the morphology of the HI density field, simulated using the `inside-out' model of reionization. In this paper we use the Largest Cluster Statistics (LCS)  to study  the percolation transition. Concerning the neutral hydrogen we find 
LCS $\approx 1$ through the entire redshift range which we have considered, $7 \lesssim z \lesssim 13$. This informs us 
that there is a single large neutral region which persists as reionization proceeds. This region percolates through the entire simulation box spanning from one face of the simulation volume to the other and formally has infinite volume.  

On the other hand, concerning the ionized regions, we find that the LCS has a small value during the early stages of reionization.  As reionization proceeds we find the onset of a transition, the percolation transition, beyond which the LCS increases sharply to attain a value $\approx 1$ which is maintained through the subsequent stages of reionization. 
The percolation transition in the ionized regions takes place at the critical neutral fraction $\xhi^C \approx 0.728$, when almost $12\%$ simulation volume is filled by the ionized hydrogen (HII), i.e. the critical filling factor $\ffc^C \approx 0.12$. These results agree well with the previous findings of \cite{Iliev2006,Chardin2012,Furlanetto2016}.  
After percolation most of the ionized volume is rapidly filled by an enormous (formally infinite) region. In the vicinity of percolation, the ionized regions follow a power law distribution for a large interval in volume.; $dn/dV \propto V^\tau$ where $\tau \approx -1.85$. This is consistent with the results of \cite{Furlanetto2016}, who find $\tau \approx -2$ for the HI fields simulated using the 21CMFAST code.

The study of Shapefinders in the vicinity of percolation reveals that, as the largest ionized region grows with reionization, its Minkowski functionals increase but their ratios, the first two Shapefinders -- thickness ($T$) and breadth ($B$) -- do not increase much. However the third Shapefinder -- length ($L$) -- increases almost linearly with volume, $L \propto V^{0.841}$. Consequently $L \gg B \simeq T$ for the largest ionized region. The product of thickness and breadth, $T \times B$, provides a measure of `cross-section' of a filament-like region. We find that the largest ionized region possesses a characteristic cross-section of $\sim 7$ Mpc$^2$ which does not vary much near the percolation transition, while the length of this region increases abruptly.  This makes the largest ionized region become very filamentary at the onset of percolation. As the largest ionized region grows, its genus increases, i.e. more tunnels pass through it. Hence the shape and the topology of the largest ionized region becomes more complex with time. 

We also study the Shapefinders of all ionized regions at various stages of reionization. We find that, at a fixed redshift, larger ionized regions are in general more filamentary and their cross-sections increase more slowly with their volumes compared to the increase in their lengths. As more large ionized regions start to appear near the percolation transition this feature becomes more pronounced.
In addition, the genus value is higher for larger ionized regions which is suggestive of their being multi-connected with complex topology. This could be because larger ionized regions grow via the merging of many filament-like smaller ones. 

As the first in a series of papers, meant to explore the shape statistics of the reionization field, this work investigates the topology and morphology of ionized bubbles as they evolve during reionization. The morphology, studied using percolation, Minkowski functionals and Shapefinders, is richer in information than more conventional probes of reionization, e.g., the two-point correlation function. In a companion paper (\cite{bag2018b}), we shall study the morphology of HI overdense and underdense excursion sets using similar tools. We also plan to include other models of reionization in our analysis and compare the topology and morphology of HI density fields simulated using these models. We also wish to extend our analysis to understand whether the data from upcoming low-frequency interferometers, such as SKA, HERA, can be used for calculating the Shapefinders. This would involve computing the Shapefinders in the presence of instrument noise and astrophysical foregrounds.


\section*{Acknowledgement}
The authors would like to acknowledge useful discussions with Tirthankar Roy Choudhury, Santanu Das, Aseem Paranjape and Ajay Vibhute. S.B. thanks the Council of Scientific and Industrial Research (CSIR), India, for financial support as senior research fellow. The HI simulations, used in this work, were done at the computational facilities at the Centre for Theoretical Studies, IIT Kharagpur, India. The numerical computations, related to percolation and shape analyses, were carried out using high performance computation (HPC) facilities at IUCAA, Pune, India.


\bsp	
\label{lastpage}
\end{document}